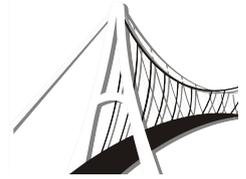

# Optimal Design of Redundant Structural Systems: fundamentals

**André T. Beck**

# Optimal Design of Redundant Structural Systems: Fundamentals

**André T. Beck,** atbeck@sc.usp.br
Department of Structural Engineering, University of São Paulo
Av. Trabalhador São-carlense, 400, 13566-590 São Carlos, SP, Brazil

**ABSTRACT**

In spite of extended recent interest in System Reliability-Based Design Optimization (System RBDO) and life-cycle cost or Risk Optimization (RO), there is a lack of published studies on optimal design of redundant hyperstatic systems with objective consideration of a) progressive collapse and b) the impact of epistemic uncertainties. This paper investigates the fundamental aspects of the problem, by addressing the optimal design of simple two-bar active and passive redundant systems. Progressive collapse is objectively addressed, differentiating consequences of direct collapse of statically determinate structures, and progressive collapse of redundant, statically indeterminate structures. A comprehensive study is performed, considering material post-failure behavior (fragile-ductile), strength correlation, dynamic amplification factors in load re-distribution, and material strength ratios. Physical uncertainty in material strengths and loads is considered. However, it is well known that reliability of a structural system also depends on nonstructural factors, or factors beyond structural design, such as unanticipated loading, manufacturing quality, quality of workmanship and human errors. These factors can be taken into account in an encompassing risk analysis, which accounts for physical and epistemic uncertainties, and which contributes a fixed latent failure probability to the structural optimization problem. Results presented herein show that the latent failure probability is the single most important parameter in determining optimal solutions, in Systems RBDO and in RO solutions. When the latent failure probability is smaller than target (System RBDO) or optimal (RO) failure probabilities, there is an equivalence between redundant and non-redundant (hyperstatic and isostatic) designs. However, when the latent reliability is smaller than target or optimal reliabilities, optimal designs become necessarily redundant (hyperstatic); as the only way to make system reliability larger than the latent reliability is by making structural systems redundant. This result is widely known in context of system reliability, but has been overlooked in past studies involving structural System RBDO and Risk Optimization.

1. **INTRODUCTION**

Until today, basic structural design is made on a member-by-member basis. This includes design of redundant, statically indeterminate structures, which are known to be more reliable than non-redundant or statically determinate structures. Reliability-based calibration of partial safety factors of modern design codes is made at member level. To the best of our knowledge, there are no practical procedures or equivalent partial factors that can be used to design redundant systems, objectively accounting for their greater reliability.

Optimal design of structural systems considering uncertainties has been object of extensive research in recent years [1-8]. Some of the early seminal papers on structural reliability theory already advocated for the use of reliability measures to drive optimal structural design [9-17]. Reliability-Based Design Optimization (RBDO) addresses minimization of an objective function involving material and/or construction costs, subject to a reliability constraint for each failure mode [5-7, 18]. RBDO is a straightforward extension from deterministic optimization, where deterministic design constraints are replaced by probabilistic constraints. System RBDO uses a single system reliability design constraint, allowing competition between failure modes [7, 19-27]. System RBDO has no deterministic equivalent, since the profession never managed to develop equivalent system safety factors. Life-cycle cost or Risk Optimization (RO) also allows competition between failure modes, but by accounting for expected consequences of failure; objectively addressing the compromise between safety and economy in structural design [7-8, 28-43, 74]. In a recent and extensive literature review [1-43], the author did not find a single reference on optimal design of redundant (hyperstatic) systems, with objective consideration of progressive collapse (see also over ninety references in Chapter 11 of Ref. [45]). Hence, optimal design of redundant systems accounting for progressive collapse is a widely open field of research.

This paper explores the fundamentals of optimal design of redundant structural systems, objectively considering progressive collapse and the effects of epistemic uncertainties. The study object is the simplest possible redundant structural system of two bars under tensile loading. Albeit simple, it reveals the influence of the most relevant variables: type of redundancy (active or passive), type of material post-failure behavior (fragile or ductile), dynamic amplification in load redistribution, and correlation between material strengths. Two unique features of the study are: differentiating consequences of collapse of redundant systems, occurring in a progressive manner (with warning), and collapses of statically determinate systems, which occur directly; and acknowledging the fundamental influence of epistemic uncertainties in optimal structural design.

This paper does not address practical aspects of design for progressive collapse, such as binding techniques, compartmentalization, structural fuses, catenary actions, etc.; it does, however, provide a framework for the probabilistic investigation of the effectiveness of such measures.

## 2. RISK ANALYSIS AND EPISTEMIC UNCERTAINTIES IN STRUCTURAL ENGINEERING

### 2.1 Risk analysis

Engineering models are approximate representations of reality. Classical models in structural engineering are derived from solid mechanics principles, and assume perfectly known material properties, loads and boundary conditions.

When dealing with model representations and modelling assumptions, different types of uncertainties need to be taken into account [44-50]. This includes: a) objective, aleatory uncertainties in structural loads, strengths of materials and member dimensions; b) subjective, epistemic uncertainties related to the amount of knowledge or information about the structural system being modelled; and c) human errors, arguably classified as aleatory or epistemic [46, 75-78]. Objective aleatory uncertainties can be quantified in terms of probabilities, with uncertain parameters modelled as random variables, random processes or random fields. Classic structural reliability analysis addresses objective uncertainties, quantified using probabilities.

Sub-modelling is employed to decompose complex engineering problems into smaller tractable units. For instance, risk analysis of a structural system considers several elements of subjective nature, and yields, among other things, potential loading scenarios (Figure 1). Risk analysis includes many nonstructural factors, like: the threats to a given structure, considering its intended use, intended public and surroundings; the quality of workmanship, especially in case of large engineering projects; human errors in construction, controls during design and execution stages, operational abuse; subjective judgement and expert opinions [71, 72], etc. The above factors include both aleatory and epistemic types of uncertainties. The threats, for instance, can be natural loading events like wind, snow, earthquakes, which can be described using probabilities; or exceptional loading arising from explosions, traffic accidents or terrorist attacks, whose quantification is more subjective. Risk analysis usually combines the above elements using failure and event trees, as illustrated in Figure 1, and results in potential loading scenarios, and in the probabilities of occurrence of each loading scenario.

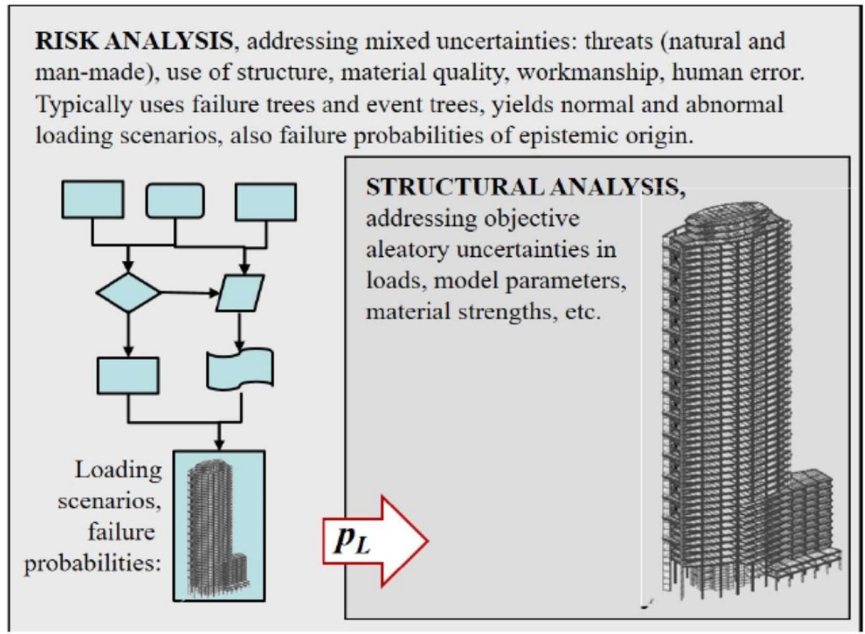

Figure 1: Relationship between risk analysis and structural analysis (with elements from [79]).

As an example, consider the case of progressive collapse due to abnormal loading. In a risk analysis, the probability of occurrence of an abnormal loading event is evaluated as $p_{ALE}$. The probability of column loss ($CL$), given the abnormal loading event, is $P[CL|ALE]$. The unconditional probability of column loss becomes $p_{CL} = P[CL|ALE]p_{ALE}$.

For a blast loading event, for instance, Shi and Stewart [80] show that collapse of a reinforced concrete column varies with size and distance of the charge; two variables that are very subjective to quantify. Similarly, the impact load resulting from a traffic accident depends strongly on mass and speed of the impacting vehicle. Due to the above epistemic uncertainties, among other reasons, design guidelines for sensitive buildings [51-55] prescribe requirements for alternate load paths, following discretionary element removal, and making designs robust w.r.t. column loss scenarios. In view of the above discussion, these design guidelines are equivalent to assuming $p_{CL} = 1$.

In a broad risk analysis addressing terrorist attacks, Stewart [81] reaches the conclusion that it may not be cost-effective to protect all US public buildings using the conservative column loss scenario.

A similar study [82] addressing performance, protection and control measures for iconic bridges shows that it may not be cost-effective to require effective performance and protection against terrorist attacks by improvised explosive devices, when the threat probability is below around $p_{ALE} \cong 10^{-4}$. In the illustrative example addressed herein, not considering the column loss scenario is equivalent to assuming $p_{CL} = 0$.

In between the two extreme scenarios of assuming $p_{CL} = 1$ and $p_{CL} = 0$ for the column loss example, in this paper we assume that one result of a risk analysis may be the failure probability for each element of the structure ($0 \leq p_{CL} \leq 1$). This failure probability is evaluated from physical and epistemic uncertainties not directly related to the structural analysis. From the point of view of the structural analysis or the optimization problem, this probability is given, and it is fixed. In order to maintain generality, this is called the **latent failure probability** ($p_L$). Consideration of such a latent failure probability is a way of acknowledging that the reliability of structural systems does not depend only on structural factors, like the number of elements or their strength: it also depends on non-structural factors like workmanship, human errors in design and operation, operational abuse, failure due to unanticipated loads, accidental loads, terrorist attacks, etc. The latent failure probability is a constant, independent of usual design loadings, the structural redundancy or the actual strength of structural elements. Depending on nature and following the risk analysis, this could be a constant member failure probability, or a constant (sub)system failure probability. Here, constant refers to the structural design problem, not to the source or the actual member. Clearly, system reliability is a function of such latent failure probability.

The impact of latent failure probabilities on the structural system also depends on the enveloping risk analysis (Figure 1). Building on the column loss example above, the column loss probability due to terrorist attack with improvised explosive devices will be greater for columns at the public entrance of a building, than for other internal columns. The probability of column loss due to traffic accident is greater for a corner column exposed to two lanes of traffic, than for other internal columns of a building. The probability of column loss is smaller for a bridge column containing a physical barrier for protection.

Unanticipated loads (of phenomenological origin, for instance) could affect the reliability at system level, or affect a group of members. Examples are the failures of Tacoma Narrows bridge and the World Trade Center, where unanticipated loads affected a large group of members. Structural compartmentalization could be used to assign different and independent $p_L$ to different parts of the structure.

The latent failure probability is evaluated from physical and epistemic uncertainties not directly related to the structural analysis. In contrast, conventional structural reliability analysis only addresses objective probabilities, using well defined probabilistic models. Because of this difference, we refer to $p_L$ as a general way of addressing epistemic uncertainties in structural reliability analysis and, more specifically, in reliability-based optimization. The latent failure probability concept explored herein can be seen as complementary to more objective approaches to address modelling uncertainties in structural design, such as those described in [83-86].

The concept of latent failure probability proposed herein has strong similarities with the epistemic reliability index proposed by Ito et al. [64], in a context of design optimization. The analysis in [64], however, specifically addresses sampling uncertainties, whereas the latent failure probability proposed herein is given a much broader meaning. For the particular case of sampling uncertainties, the authors [64] are able to show that a conservative reliability index is obtained as the sum of the epistemic and aleatory reliability indexes; and that the epistemic reliability index is independent of the design variables. By similarity, the latent failure probability described herein is also independent of the (structural) design variables. Instead of adding the latent failure probability to the failure probability evaluated from aleatory variables, as in [64], we propose it to be included in the structural system modeling.

Risk analysis considering epistemic uncertainties often involves possibilistic representations of uncertainties, using intervals, fuzzy numbers, fuzzy probabilities or imprecise probabilities [56-58]. This is the case when subjective judgement and expert opinion [71, 72] need to be accounted for. This may result in interval or fuzzy load cases, or in interval or fuzzy latent failure probabilities. These cases are out of scope for this paper.

## 2.2 Epistemic uncertainties in structural reliability and optimization

Epistemic uncertainties are also often included in structural reliability analysis. When epistemic uncertainties are quantified probabilistically, failure probabilities are interpreted as point-estimates [48]. Usually, however, epistemic uncertainties are modelled using intervals, fuzzy numbers, fuzzy probabilities or imprecise probabilities [56-58]. This leads to interval or fuzzy failure probabilities. The simultaneous consideration of interval or fuzzy variables, and random variables, leads to nested uncertainty representations, requiring failure probabilities to be evaluated repeatedly [60].

Epistemic uncertainties have also been considered in the context of reliability-based design optimization [60-67] and risk optimization [68-70]. Approaches specifically addressing insufficient information [61-64] lead to compromise solutions between reducing the objective function and increasing confidence on constraint feasibility. Approaches addressing more general types of epistemic uncertainties either consider conservative failure probability estimates [60, 64, 65], or lead to nested loops for handling interval/fuzzy and random variables. In the later case, the two nested uncertainty modelling loops become nested with the optimization problem, leading to three nested computational loops. Most of the papers in this subject are directed towards decoupling the analysis loops, making the computations tractable [59, 60, 67]. The interested reader is referred to Jiang et al. [60] for a comprehensive review on the subject.

In this paper, we propose an alternative and/or complementary way of considering the effect of epistemic uncertainties in structural reliability analysis or in reliability-based optimization. We propose epistemic uncertainties to be handled in the encompassing risk analysis, and that their effect on (the optimal design of) structural systems be reduced to a given, fixed latent failure probability. We acknowledge that this may not be possible for all problems; herein, we address those cases in which this is possible. In this sense, we are not trying to compete with alternative ways of handling epistemic uncertainties [56-60, 66-70, 83-86]. We do observe, however, that in those cases where the proposed alternative is feasible, the proposed latent failure probability approach has some advantages: a) it avoids the nested loops for handling epistemic and aleatory uncertainties; b) effects of the two sources of uncertainty are separated and easier to understand; and c) the latent failure probability is evaluated (estimated) only once, and remains constant during the design process. Notice the similarities with the properties of the epistemic reliability index in [64].

As observed above, in this manuscript we do not propose to compete with alternative ways of addressing epistemic uncertainties in structural reliability and reliability-based optimization problems [56-60, 66-70]. We include this sub-section only because we foresee a generalization, beyond the scope of this manuscript, to handling of epistemic uncertainties in reliability-based optimization.

The main objective of this manuscript, in this regard, is to evaluate the impact of such latent failure probabilities (due to epistemic uncertainty) in optimal design of redundant structural systems. In spite of the tremendous impact (to be shown), there is no evidence of the latent failure probability concept in the published structural reliability or optimization literature. An extensive literature review on the subject [1-43, 59-70] revealed a single paper with an equivalent idea [64], yet focused on sampling uncertainties. In a similar way, the actual risk analysis is out of scope of this manuscript, which addresses the impact on optimal system reliability of a given latent failure probability $p_L$.

## 3. STUDY OBJECT AND BASIC FORMULATION

### 3.1 Study object

This paper addresses optimal design of simple redundant systems, such as those illustrated in Figure 2. The systems are formed by two bars of different materials, identified with numerals 1 and 2, which share the tensile load *P*.

The study object is simple and academic; but it serves the purpose of evaluating the impact of epistemic uncertainties in optimal design of redundant structural systems. Extensions to multi-bay multi-story RC buildings, with more explicit modelling of progressive failure, will be the object of future studies.

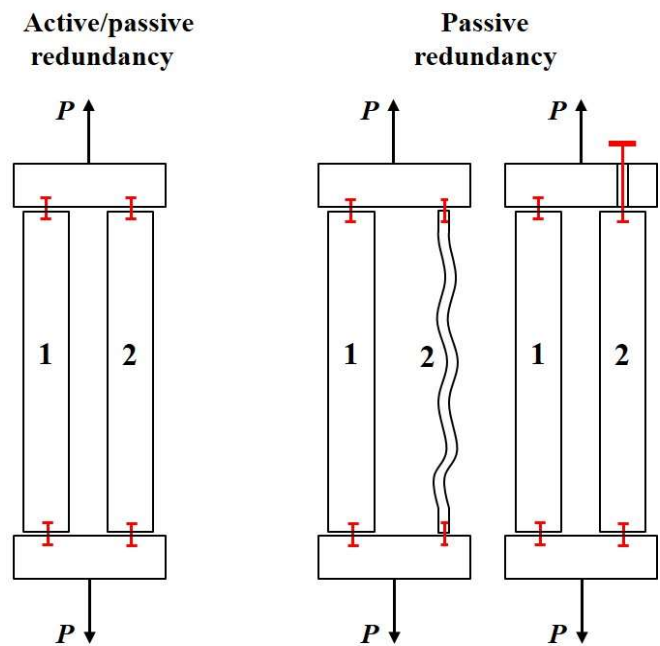

Figure 2: Statically indeterminate systems with active-passive (left)
and passive (right, two structures) redundancies.

The cross-section areas of the two bars are the design variables of the optimization problem. These are determined independently. We start with a statically indeterminate two bar system, and let the optimizer decide whether a second bar is necessary or not. When the second bar is kept, we don´t know *a priory* if its cross-section area is sufficient to sustain load *P*. Hence, it is up to the optimizer to decide whether redundancy will mean just a statically indeterminate system, or if redundancy means that the secondary element has sufficient capacity to absorb the load in case of failure of the primary element.

When the structure is statically indeterminate and the two bars share the loading, we have a system with active redundancy (Figure 2, left). If the secondary member has sufficient capacity to absorb the whole load *P*, in case of failure of the primary member, then the reserve strength of the secondary member is mobilized in a passive manner. In this case, the redundant system illustrated in Figure 2 (left) actually exhibits active and passive redundancies. Therefore, from this point on, this is called active-passive redundancy in this paper.

The two systems illustrated in Figure 2 right exhibit passive redundancy, because the second element is disconnected, and does not take any loading, unless element 1 fails. The strength of element 2 is only mobilized in case of failure of element 1. Formulation and results for passive systems are presented in the appendix.

## 3.2 Material model

Progressive failure, and reserve strength of "failed" members depends on material behavior. In this study, a simple fragile-ductile material model is considered, as shown in Figure 3. The rupture or yielding stress is $S$, and the post-failure strength is given by $\eta S$. For $\eta = 0$ we have elastic-fragile material; for $\eta = 1$ the material is elastic-perfectly plastic (ductile). Intermediate behavior is obtained for $0 < \eta < 1$.

In order to make developments as analytical and linear as possible, material strengths (and loading, for this matter) are modelled as Gaussian random variables. Rupture or yielding stresses are given by:

$$S_1 \sim N(\mu_1, \sigma_1), \\ S_2 \sim N(\mu_2, \sigma_2), \tag{1}$$

where the subindex $()_i$ indicates bar number. Coefficient of variation (COV) of material strength is given by $\delta_i = \sigma_i/\mu_i$. Moreover, correlation between material strengths is given by $\rho_{12} \in [0,1)$.

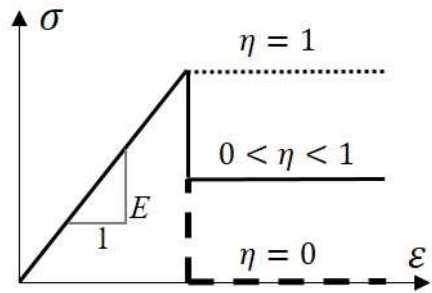

Figure 3: Fragile-ductile material model.

### 3.3 Load model

In order to investigate progressive failure of a two-bar parallel system, a sequence of two load applications needs to be considered. In this paper, loading is modelled as an independent sequence of two Gaussian load pulses ($P_1 = P_2 = P$), with identical intensity:

$$P_1 \sim N(\mu_P, \sigma_P); \; P_2 \sim N(\mu_P, \sigma_P). \tag{2}$$

Each load pulse represents the maximum load that would be sustained during two periods of operation.

In case of failure of a primary member, dynamic load amplification may occur. To account for dynamic load amplification, an impact factor $f_i$ is considered:

$$f_i = \frac{P_R}{P} \geq 1, \tag{3}$$

where $P_R$ is the load in the secondary member immediately after redistribution. The second load application ($P_2$) is assumed to occur under static conditions. In the following, dynamic load amplification factors $f_i = \{1, 1.3\}$ are considered. Load amplification is particularly relevant in failure of brittle members; nevertheless, the above values are employed for brittle and ductile members in the sequence.

### 3.4 Problem formulation

Let **X** and **d** be vectors of structural system parameters. Vector **X** contains random variables, such as member dimensions and geometry, strength of materials, loads and model error variables. Vector **X** represent the objective uncertainties, which can be modelled using probabilities (random variables). Vector **d** contains design variables whose values are to be determined, in order to maximize performance of the system, or in order to minimize weight, cost, etc. Typical variables in vector **d** are nominal member dimensions, partial safety factors, reinforcement ratio, design life, parameters of inspection and maintenance programs, etc.

To keep the development as simple as possible, in this paper vector **X** contains only the rupture or yielding stress and the loads, such that: $\mathbf{X} = \{S_1, S_2, P_1, P_2\}$. Vector **d** contains the partial load factors to be used in design of the redundant systems: $\mathbf{d} = \{\lambda_1, \lambda_2\}$, which are introduced shortly.

The existence of uncertainty implies the possibility of structural failure. The boundary between safe and failure domains is given by limit state functions $g_i(\mathbf{d}, \mathbf{X}) = 0$, written for each failure mode, and/or for each member of the structure, such that:

$$\Omega_{fi}(\mathbf{d}) = \{\mathbf{x}|g_i(\mathbf{d}, \mathbf{X}) \leq 0\}, \tag{4}$$

$$\Omega_{si}(\mathbf{d}) = \{\mathbf{x}|g_i(\mathbf{d}, \mathbf{X}) > 0\}, \quad i = 1, \ldots, n_{LS},$$

where $\Omega_{fi}(\mathbf{d})$ is the failure domain, $\Omega_{si}(\mathbf{d})$ is the survival domain, and $n_{LS}$ is the number of limit state functions. Limit states describe primary and conditional member failures, as well as simultaneous system failure. System failure resulting from progressive collapse is characterized by specific combinations of primary and conditional failures, as shown in the sequence.

Probabilities of primary and conditional member failures are given generically by:

$$p_{fi} = P[\mathbf{X} \in \Omega_{fi}] = \int_{\Omega_{fi}} f_\mathbf{X}(\mathbf{x})\, d\mathbf{x}, \tag{5}$$

where $P[\ ]$ is the probability operator and $f_\mathbf{X}(\mathbf{x})$ is the joint density function of the random variable vector. In this manuscript, limit state functions are linear, and random variables have Gaussian distribution, hence Eq. (5) has analytical closed form solution [44, 45]:

$$p_{fi} = \Phi[-\beta_i], \quad \text{with} \quad \beta_i = \frac{E[g_i(\mathbf{d}, \mathbf{X})]}{\sqrt{Var[g_i(\mathbf{d}, \mathbf{X})]}}, \tag{6}$$

where $\beta_i$ is the reliability index, $E[\ ]$ is the expected value operator, and $Var[\ ]$ is the variance operator.

For the progressive collapse of redundant, statically indeterminate structural systems, the limit state function is written generically as:

$$\Omega_f(\mathbf{d}) = \{\mathbf{x}|\ \cup_k \cap_{i \in C_k} g_i(\mathbf{d}, \mathbf{X}) \leq 0\}, \tag{7}$$

where $\Omega_f(\mathbf{d})$ represents the compound system failure domain, $C_k$ is the index set listing components of the $k^{th}$ cut-set, such that series, parallel and combined cut-set systems can be represented. Let $p_{fsys}$ represents the system failure probability, such that:

$$p_{fsys} = P[\mathbf{X} \in \Omega_f] = \int_{\cup_k \cap_{i \in C_k} g_i(\mathbf{d}, \mathbf{X}) \leq 0} f_\mathbf{X}(\mathbf{x})\, d\mathbf{x}\ . \tag{8}$$

### 3.5 Latent failure probability of the parallel system

As discussed in Section 2, the reliability of structural systems does not depend only on structural factors. Human errors in design, workmanship and operation, unanticipated loading and phenomenological uncertainties may lead to failures which are usually not considered in limit state functions (like Eqs. 4 and 7), but which can be evaluated from an encompassing risk analysis (Figure 1).

This paper addresses optimal design of the cross-section of the structural elements in Figure 2, for the normal load case described in Section 3.3. However, it is assumed that an encompassing risk analysis identified some abnormal loading conditions of very large intensity but very low probability of occurrence; it was further found that it is not economical to design individual elements to withstand such abnormal loading [80, 81]. The single element failure probability given in risk analysis is $p_L$. The same $p_L$ is valid for both elements.

To maintain generality, an alternative justification for the same $p_L$ comes from reliability of the connections, as illustrated in Figure 2. Technically, strength of the connections could be part of the structural design problem; but connections also fail due to non-structural causes like poor workmanship, manufacturing defects or assembly mistakes, which also vary according to connection type (welded, bolted, glued, etc.) and to experience of the construction crew. The probability of failure resulting from risk analysis could also be due to connection reliability ($p_{CON}$) and unanticipated loading ($p_{UL}$), such that $p_L = p_{CON} + p_{UL}$.

As the structural system considered herein is very simple, the interpretation of $p_L$ as resulting from abnormal loading (and producing a "column loss" scenario) is not straightforward; therefore, from this point on, it is considered that $p_L$ results from connection failure due to non-structural causes. The main results presented in the sequence; however, are also valid for a situation of "column loss due to abnormal loading".

Let $c_1$ be the event "failure of connection to element one", and $c_2$ be the event "failure of connection to element two", with probabilities:

$$P[c_1] = p_{L1}; \ P[c_2] = p_{L2}. \tag{9}$$

where $p_L$ stands for latent failure probability. In order to avoid excessive growth of the failure tree, it is assumed that connection failure can only occur at the first time the connection is loaded (the first load serves as proof load).

Following Eq. (6), the latent failure probability can also be written as a latent reliability index $\beta_L = -\Phi^{-1}[p_L]$. This allows straightforward comparison with the target reliability index $\beta_T$ in reliability-based optimization (RBDO), and with the optimal reliability index $\beta^*$ of risk-based optimization (both to be described).

## 4. DESIGN AND RELIABILITY OF SYSTEM WITH ACTIVE-PASSIVE REDUNDANCY

### 4.1 Load sharing

Consider the two-bar statically indeterminate parallel system illustrated in Figure 2. In this type of (conventional) redundant system, the two bars share the workload, until one or both of them fail. If there is passive or standby redundancy, the remaining bar will absorb the total load, as commented in Section 3.1. If the displacement $u$ of both bars is the same, then the fraction of load P carried by the $i^{th}$ bar is:

$$N_i = k_i u = P \frac{k_i}{(k_1+k_2)}, \tag{10}$$

where $k_i$ is the stiffness of the $i^{th}$ bar, given by: $k_i = E_i a_i / L_i$, with $E_i$ the elasticity modulus, $a_i$ the cross-section area, and $L_i$ the length of each bar. If elasticity modulus and length are the same, the load fraction reduces to:

$$N_i = P \frac{a_i}{(a_1+a_2)}. \tag{11}$$

### 4.2 Usual design

Usual design of redundant statically indeterminate systems is made on a member-by-member basis, without differentiating for number of parallel elements. Moreover, load sharing is decided beforehand, i.e., on a basis of analyzing influence areas for loads. For the simple redundant system illustrated in Figure 2, for instance, one has:

$$a_1 \mu_1 + a_2 \mu_2 = \lambda_E \mu_P, \tag{12}$$

where $\lambda_E$ is a conventional load (safety) factor. For proportional load sharing, one has $a_1 \mu_1 = a_2 \mu_2$, which inserted in Eq. (12) leads to:

$$a_i(\lambda_E) = \frac{\lambda_E \mu_P}{2\mu_i}, \tag{13}$$

### 4.3 Design with independent dimensioning of each member

In order to address the optimal design of active-passive redundant systems, independent proportioning of each element is considered herein:

$$\begin{aligned} R_D &\geq S_D \\ a_1 \mu_1 &= \lambda_1 \mu_P \\ a_1(\lambda_1) &= \frac{\lambda_1 \mu_P}{\mu_1}, \end{aligned} \qquad (14)$$

and

$$\begin{aligned} R_D &\geq S_D \\ a_2 \mu_2 &= \lambda_2 \mu_P \\ a_2(\lambda_2) &= \frac{\lambda_2 \mu_P}{\mu_2}, \end{aligned} \qquad (15)$$

where $\lambda_i \geq 0, i = \{1, 2\}$, is the partial load factor for element $i$. The two-bar active-passive system is symmetrical; hence, any of the two bars could fail first, the other bar becoming the redundant element. Each load factor is allowed to be equal to zero (but not both); hence, the result of system optimization may be that a redundant element is not necessary.

### 4.4 System reliability

The failure pattern of active-passive systems has both elements sharing the load from the beginning. Figure 4 illustrates the event tree for two load applications, where $F_i$ denotes the event "failure of bar $i$", and $c_i$ denotes the event "failure of connection to bar $i$". Overhead bars represent the complementary event. Failure consequences are discussed later.

When the structure is first loaded, both bars are active; hence, both connections can fail upon first loading. This leads to four possible events, w.r.t. connection failures:

$$\{ (c_1, c_2), (\bar{c}_1, c_2), (c_1, \bar{c}_2), (\bar{c}_1, \bar{c}_2) \} \qquad (16)$$

If both connections fail $(c_1, c_2)$, we have direct collapse (DC). If only one connection fails, the conditional failure events illustrated in Figure 4 become possible. For these conditional events, residual strength of the failed bar is not considered ($\eta = 0$), as the failed bar is actually not connected. Only if both connections survive the first load application, the upper-branch chain of events ilustrated in Figure 4 becomes possible.

Due to independent dimensioning (Eqs. 14 and 15), any bar can fail first. The limit state function for primary failure of the $i^{\text{th}}$ bar ($F_i$) is given by:

$$g_i(\mathbf{d}, \mathbf{X}) = a_i S_i - N_i = a_i S_i - P \frac{a_i}{(a_1 + a_2)} = (a_1 + a_2) S_i - P = 0 \qquad (17)$$

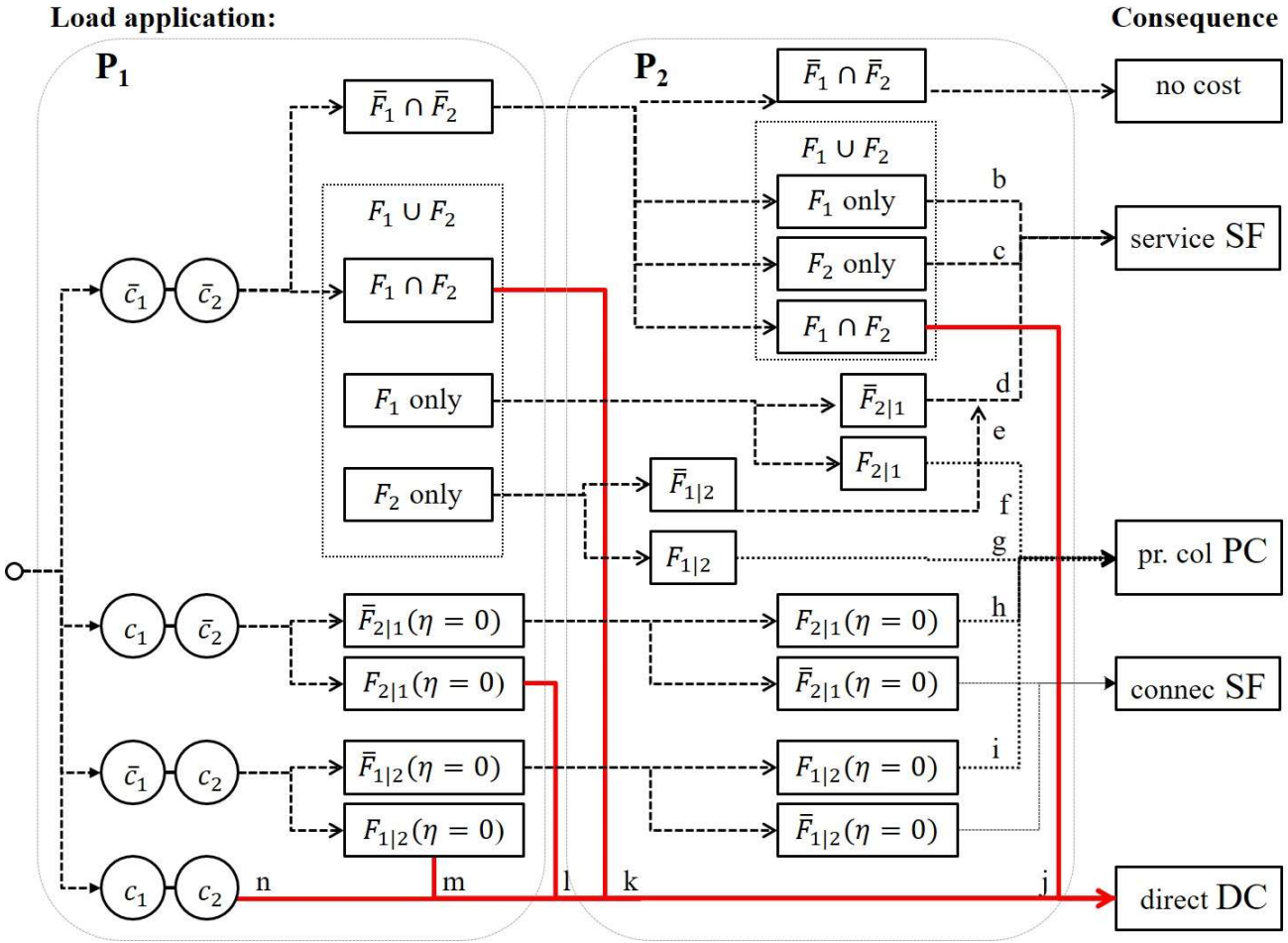

Figure 4: Event tree for active-passive system submitted to two load applications.

The reliability index for primary member failure is:

$$\beta_i(\mathbf{d}) = \frac{(a_1+a_2)\mu_i - \mu_P}{\sqrt{(a_1+a_2)^2 \sigma_i^2 + \sigma_P^2}}. \tag{18}$$

The limit state function for conditional failure of a second element, given failure of the primary element (event $F_{2|1}$), is:

$$g_{2|1}(\mathbf{d},\mathbf{X}) = a_2 S_2 + \eta a_1 S_1 - f_i P \tag{19}$$

The reliability index for conditional failure is given by:

$$\beta_{2|1}(\mathbf{d}) = \frac{a_2 \mu_2 + \eta a_1 \mu_1 - f_i \mu_P}{\sqrt{a_2^2 \sigma_2^2 + \eta^2 a_1^2 \sigma_1^2 + \eta a_1 a_2 \sigma_1 \sigma_2 \rho_{12} + f_i^2 \sigma_P^2}}. \tag{20}$$

Progressive failure of a second element may occur in the first load application, following failure of the connection to the primary element; or in the second load application, following failure of the primary element or its connection, in the first load application. For the second load application, there is no dynamic amplification, and $f_i = 1$ in Eqs. (19) and (20).

As they are actively engaged (active redundancy), both elements may fail simultaneously. The limit state function for joint failure of the two bars (event $F_{1 \cap 2}$), is:

$$g_{1 \cap 2}(\mathbf{d}, \mathbf{X}) = a_1 S_1 + a_2 S_2 - P \tag{21}$$

The reliability index for the joint failure event is:

$$\beta_{1 \cap 2}(\mathbf{d}) = \frac{a_2 \mu_2 + a_1 \mu_1 - \mu_P}{\sqrt{a_1^2 \sigma_1^2 + a_2^2 \sigma_2^2 + a_1 a_2 \sigma_1 \sigma_2 \rho_{12} + \sigma_P^2}} . \tag{22}$$

The event in dashed lines, in Figure 4, corresponds to failure of bars 1 and/or 2: $F_1 \cup F_2$. The complementary event to $F_1 \cup F_2$ is $\bar{F}_1 \cap \bar{F}_2$, as stated in Figure 4. The union probability is written as:

$$P[F_1 \cup F_2] = P[F_1] + P[F_2] - P[F_1 \cap F_2] \tag{23}$$

By adding and substracting the term $P[F_1 \cap F_2]$ :

$$P[F_1 \cup F_2] = P[F_1] - P[F_1 \cap F_2]$$
$$+P[F_2] - P[F_1 \cap F_2]$$
$$+P[F_1 \cap F_2] \tag{24}$$

we obtain, in the first line of Eq. (24), the probability that only bar 1 fails :

$$P[F_1 \text{ only}] = P[F_1] - P[F_1 \cap F_2] = \Phi[-\beta_1] - \Phi[-\beta_{1 \cap 2}] . \tag{25}$$

In the same way, the second line in Eq. (24) is the probability that only bar 2 fails.

If the connections do not fail, there are tree mutually exclusive ways in which system failure can occur: progressive failure of bars in order 1-2, progressive failure in order 2-1, or simultaneous failure of both bars. Hence, following the event three in Figure 4, the system failure probability is obtained as:

$$p_{fsys}(\mathbf{d}) = (1 - p_{L1})(1 - p_{L2})\{(P[F_1 \text{ only}] )\Phi[-\beta_{2|1}] \tag{f}$$

$$+(P[F_2 \text{ only}])\Phi\left[-\beta_{1|2}\right] + \Phi[-\beta_{1 \cap 2}]\} \tag{g,k}$$

$$+(p_{L1})(1 - p_{L2})\left(1 - \Phi\left[-\beta_{2|1}(\eta = 0)\right]\right)\Phi\left[-\beta_{2|1}(\eta = 0, f = 1)\right] \tag{h}$$

$$+(p_{L1})(1 - p_{L2})\Phi\left[-\beta_{2|1}(\eta = 0)\right] \tag{l}$$

$$+(1 - p_{L1})(p_{L2})\left(1 - \Phi\left[-\beta_{1|2}(\eta = 0)\right]\right)\Phi\left[-\beta_{1|2}(\eta = 0, f = 1)\right] \tag{i}$$

$$+(1-p_{L1})(p_{L2})\Phi\left[-\beta_{1|2}(\eta=0)\right] \quad \text{(m)}$$

$$+p_{L1}p_{L2} \quad \text{(n)}$$

$$(26)$$

where the term $\beta_{1|2}$ is evaluated from Eq. (20), with a proper change of indexes. In Eq. (26), the letters identify each line, according to the failure tree in Figure 4. Following Eq. (6), the system reliability index can be evaluated from system failure probability as $\beta_{sys} = -\Phi^{-1}[p_{fsys}]$. Eq. (26) also assumes independence between the connection failure events $c_1$ and $c_2$. More details about system reliability modelling can be found in [74].

The present academic problem was built so as to admit exact solution by the FOSM method (linear limit state functions of Gaussian random variables). In general, more elaborate reliability techniques may be required [see 44-46, for instance].

## 5. DESIGN OPTIMIZATION

From Eqs. (17) and (26), it is clear that the reliability of the redundant system, $\mathcal{R}(\mathbf{d}) = 1 - p_{fsys}(\mathbf{d})$, is a function of the partial load factors to be used in design: $\mathbf{d} = \{\lambda_1, \lambda_2\}$. Optimum design considering progressive failure is sought, with the objective of minimizing cost of materials:

$$h(\mathbf{d}) = a_1(\lambda_1)\mu_1 + a_2(\lambda_2)\mu_2 \quad (27)$$

where unit material cost per volume is simply assumed proportional to mean strength ($\mu_i, i = 1,2$). Two different approaches can be used to formulate the design problem: reliability-constrained design optimization, or optimization considering failure consequences, as follows.

### 5.1 Reliability-constrained design optimization problem (System RBDO)

In reliability-constrained design optimization, also known as System Reliability-Based Design Optimization (System RBDO) in the literature, minimization of an objective function $h(\mathbf{d})$ is sought, with constraints expressed in terms of a target (or minimum) system reliability [19-27,74]:

Find $\mathbf{d}^* = \{\lambda_1^*, \lambda_2^*\}$ which minimizes $h(\mathbf{d})$

subject to: $\mathcal{R}(\mathbf{d}) = 1 - p_{fsys}(\mathbf{d}) \geq \mathcal{R}_T$; $\mathbf{d} \in \mathcal{D}$, $\quad (28)$

where $\mathcal{R}_T$ is the target system reliability and $\mathcal{D} = \{0 \leq \lambda_1, 0 \leq \lambda_2\}$ are side constraints. Following Eq. (6), the target system reliability can also be written in terms of a target reliability index: $\beta_T = \Phi^{-1}[\mathcal{R}_T]$.

Results of RBDO are clearly dependent on the target system reliability $\beta_T$ used as design constraint. The target reliability is specified a priori by the designer, and should be based on a cost-benefit analysis.

## 5.2 Risk optimization

Risk optimization [28-43,74] increases the scope of RBDO by including the expected consequences of failure. Each failure event described in Figure 4 has different consequences. In a redundant structure, failure of a primary element can be considered as a service failure; which demands corrective maintenance. Failure of the redundant element leads to structural collapse. This may result in damage to equipment, damage to third parties, injury, death, and environmental damage. The cost of failure, or a measure of the consequences of failure, is given by $C_f$. The expected cost of failure ($C_{EF}$) is obtained by multiplying failure costs ($C_f$) by failure probabilities ($p_f$):

$$C_{EF}(\mathbf{d}) = C_f p_f(\mathbf{d}). \tag{29}$$

The cost of a primary element failure, or service failure (*SF*), is assumed proportional to the cost of materials for replacing the failed element:

$$C_{fi} = k_{SF}\mu_i a_i(\mathbf{d}) \tag{30}$$

where $k_{SF} > 1$ is a multiplication factor to account for workmanship and repair downtime.

The cost of ultimate failures accounts for damage to equipment, damage to third parties, payment of compensation for injury, death, and environmental damage, and for bad reputation. These costs are assumed constant, or independent of the design vector.

One relevant feature of risk or failure consequence-driven design optimization is the possibility to differentiate for the consequences of collapse that occurs directly, or without warning, and collapse that occurs in a progressive manner [42]. Failure of a primary element, which is not immediately followed by failure of the redundant element, serves as warning, allowing evacuation and reducing costs of compensation for injury and death. Such warning also allows preventive collapse and mitigation actions to be taken. Hence, consequences of stepped progressive collapse (*PC*) are measured by $k_{PC}$, and consequences of direct collapse (*DC*) are given by $k_{DC} > k_{PC}$.

The objective function for consequence-based optimization is obtained by adding all cost over the life-cycle of the structure. This can include costs of operation, inspection, maintenance, and cost of disposal; all of which are application-dependent. In this paper, only cost of materials, and expected costs of failure are considered.

Figure 4 illustrates failure paths and failure consequences of the active-passive two-bar redundant system. There are four paths leading to each of the consequence scenarios (*SF*, *DC*, *PC*). Two paths lead to service failure for bar 1: one with probability $(1 - \Phi(-\beta_1))(P[F_1 \text{ only}])$, and one with probability $(P[F_1 \text{ only}])(1 - \Phi[-\beta_{2|1}])$; similarly for service failure of bar 2. Half of the collapse paths are caused by connection failures.

When connections do not fail, there are two failure paths leading to progressive collapse, corresponding to failure sequences 1-2 and 2-1. For failure sequence 1-2, the probability is $(P[F_1 \text{ only}])\Phi[-\beta_{2|1}]$. When connections do not fail, there are two failure paths leading to direct collapse, corresponding to the first and second load applications, respectivelly. The probability of direct collapse is given by $P[\bar{F}_1 \cap \bar{F}_2]\Phi[\beta_{1\cap 2}] + \Phi[-\beta_{1\cap 2}]$. With these preliminaries, the objective function for active-passive redundant systems is given by:

$$h_{RO}(\mathbf{d}) = \frac{(a_1(\lambda_1)\mu_1 + a_2(\lambda_2)\mu_2)}{(a_1(1)\mu_1 + a_2(1)\mu_2)} \tag{a}$$

$$+k_{SF}\frac{a_1(\lambda_1)}{a_1(1)}(1 - p_{L1})(1 - p_{L2})(P[F_1 \text{ only}])(1 - P[F_1 \cup F_2])(1 - \Phi[-\beta_{2|1}]) \tag{b,d}$$

$$+k_{SF}\frac{a_2(\lambda_2)}{a_2(1)}(1 - p_{L1})(1 - p_{L2})(P[F_2 \text{ only}])(1 - P[F_1 \cup F_2])(1 - \Phi[-\beta_{1|2}]) \tag{c,e}$$

$$+k_{PC}(1 - p_{L1})(1 - p_{L2})(P[F_1 \text{ only}])\Phi[-\beta_{2|1}] \tag{f}$$

$$+k_{PC}(1 - p_{L1})(1 - p_{L2})(P[F_2 \text{ only}])\Phi[-\beta_{1|2}] \tag{g}$$

$$+k_{PC}(p_{L1})(1 - p_{L2})(1 - \Phi[-\beta_{2|1}(\eta = 0)])\Phi[-\beta_{2|1}(\eta = 0, f = 1)] \tag{h}$$

$$+k_{PC}(1 - p_{L1})(p_{L2})(1 - \Phi[-\beta_{1|2}(\eta = 0)])\Phi[-\beta_{1|2}(\eta = 0, f = 1)] \tag{i}$$

$$+k_{DC}(1 - p_{L1})(1 - p_{L2})\Phi[-\beta_{1\cap 2}][1 + (1 - P[F_1 \cup F_2])] \tag{j,k}$$

$$+k_{DC}(p_{L1})(1 - p_{L2})\Phi[-\beta_{2|1}(\eta = 0)] \tag{l}$$

$$+k_{DC}(1 - p_{L1})(p_{L2})\Phi[-\beta_{1|2}(\eta = 0)] \tag{m}$$

$$+k_{DC}(p_{L1})(p_{L2}) \tag{n}$$

$$\tag{31}$$

where the letters identifying each line correspond to the letters of the failure paths in Figure 4. The union term in Equation (31) is evaluated as $P[F_1 \cup F_2] = \max[0, \Phi[-\beta_1] + \Phi[-\beta_2] - \Phi[-\beta_{1\cap 2}]]$.

The failure consequence-driven design optimization problem is stated as:

Find $\mathbf{d}^* = \{\lambda_1^*, \lambda_2^*\}$  which minimizes $h_{RO}(\mathbf{d})$

subject to:  $\mathbf{d} \in \mathcal{D}$. (32)

with $\mathcal{D} = \{0 \leq \lambda_1, 0 \leq \lambda_2\}$ being side constraints.

In the above equations, failure cost multipliers $k_{SF}$, $k_{PC}$ and $k_{DC}$ can be interpreted as the actual cost of failures, or simply as parameters to balance the relative consequences of failure, w.r.t. the cost of materials for the reference structure. Typical values of these cost multipliers for conventional civil engineering structures are given in [73], which recommends a cost-benefit analysis for $k > 20$. Our structural system is simple; hence, failure consequences could be several times larger than costs of materials. In this study, the following values are used: $k_{SF} = 2$, $k_{PC} = 20$ and $k_{DC} = 100$. Specific results to be presented depend directly on these cost multipliers. However, the overall conclusions are valid for similar relative magnitudes of cost multipliers.

The risk optimization formulation addresses the safety-economy tradeoff in structural design [28-43]. The optimal reliabilities $\beta^*$ are a sub-product of the analysis. When the problem is formulated considering failure of individual members, system failure due to progressive collapse and direct system failure, all optimal reliabilities are found by risk optimization [42]. In this paper, we discuss results related to the optimal system reliability $\beta_{sys}^*$, following Eq. (26).

## 6. NUMERICAL RESULTS

### 6.1 Parametric analysis

The key parameters that characterize design of redundant systems considering progressive failure are the latent failure probability ($p_L$), dynamic load re-distribution factor ($f_i$), the correlation between material strengths ($\rho_{12}$), material behavior (fragile-ductile) and the type of redundancy (active or passive). Latent failure probabilities are considered same for both connections: $p_{L1} = p_{L2} = p_L$.

In the following, two levels of each parameter are considered: $p_L = \{0\ ;10^{-3}\}$ for System RBDO, $p_L = \{10^{-2}; 10^{-3}\}$ for RO, $f_i = \{1; 1.3\}$, $\rho_{12} = \{0; 0.9\}$, $\eta = \{0; 1\}$, combined with active-passive redundancy. Passive (standby) redundancy is addressed in the appendix. The above leads to $2^4 = 16$ combinations for each type of structural system. Further, we study different ratios between mean material strengths ($\mu_2/\mu_1$) and between strength COVs ($\delta_1/\delta_2$). Hence, in order to limit results, in risk optimization only $\rho_{12} = 0$ is considered. The latent reliability index for $p_L = 10^{-3}$ is $\beta_L = -\Phi^{-1}[10^{-3}] = 3.09$.

In the following, we discuss redundant assemblies made with two bars of brittle or ductile materials. Mixed material assemblies are not considered, but the ductile-ductile combination also describes what would be obtained for a ductile-brittle combination, if the primary failure is in the ductile material (as post-failure behavior of the second bar does not matter). In this regard, a brittle-ductile combination does not make much sense, and is also not considered.

Results in this section are evaluated for $(\mu_P = 10, \delta_P = 0.3)$ and, unless otherwise stated, $\delta_1 = \delta_2 = 0.1$. These are typical values for civil engineering structures. In the following, we consider two bars of same material ($\mu_1 = \mu_2 = 5$), and then two bars with different mean strengths but same COV: ($\mu_1 = 1, \mu_2 = 9$). Finally, the case with different mean strengths and different COVs is investigated: ($\mu_1 = 1, \mu_2 = 9, \delta_1 = 0.15, \delta_2 = 0.05$). The investigated cases are summarized in Table 1. Two reciprocal cases are also investigated, following Table 1. Results are not presented or discussed herein, as they are reciprocal to the results presented in Sections 6.2 and 6.3 (the problem is symmetric w.r.t. bars 1 and 2).

Table 1: Summary of material property cases investigated.

| Case/subsection | | $\mu_1$ | $\mu_2$ | $\delta_1$ | $\delta_2$ |
|---|---|---|---|---|---|
| **6.2** | Same mean | 5 | 5 | 0.1 | 0.1 |
| **6.3** | Different mean | 1 | 9 | 0.1 | 0.1 |
| **6.4** | Diff. mean and COV | 1 | 9 | 0.15 | 0.05 |
| - | Reciprocal to **6.3** | 9 | 1 | 0.1 | 0.1 |
| - | Reciprocal to **6.4** | 9 | 1 | 0.05 | 0.15 |

## 6.2 Results for two bars of same mean strength ($\mu_1 = \mu_2$)

**Results for System RBDO**

By using system reliability as a design constraint (System RBDO), the results illustrated in Figure 5 are obtained. Results for fragile material ($\eta = 0$) are shown at the top, and for the ductile material ($\eta = 1$) at the bottom. Results neglecting latent failure probability ($p_L = 0$) are shown left in Figure 5, and results for $p_L = 10^{-3}$ are shown right.

First, we note that the latent probability has the greatest impact in results, which superseeds any effects of material behavior (ductile/fragile), strength correlation or dynamic load re-distribution. When the latent failure probability is neglected, RBDO solutions are linear in partial factors $\lambda_1$ and $\lambda_2$. This also occurs when the target reliability index is smaller than the latent reliability ($\beta_T < 3.09$ in Figure 5, right). When RBDO solutions are linear in partial factors $\lambda_1$ and $\lambda_2$, there is a linear trade-off between these parameters. Hence, designs with same reliability are obtained with a single bar of either material, with two bars of same cross-section area, or with two bars of inversely proportional cross-section areas. However, when the target reliability is larger than the latent reliability ($\beta_T > 3.09$), single-bar solutions are no longer optimal. Two bars become necessary, as the reliability of the system can only be larger than the latent reliability if the system becomes redundant. This is an obvious result, in retrospect, which had not been demonstrated before.

When the system becomes necessarily redundant, RBDO solutions become dependent on impact factor. Further, we notice that the solutions for ductile material are more dependent on strength correlation $\rho_{12}$ for large $\beta_T$, when latent reliability is neglected or when it is much smaller than $\beta_T$.

Results obtained for RBDO serve as a hint to results for risk optimization. Specifically, the non-linear behavior related to the latent probability can be expected to affect risk optimization. Since the optimal reliability index $\beta^*$ is a by-product of risk optimization, it is expected that the risk-optimal structure will be a two-bar redundant system when $\beta^* > \beta_L$.

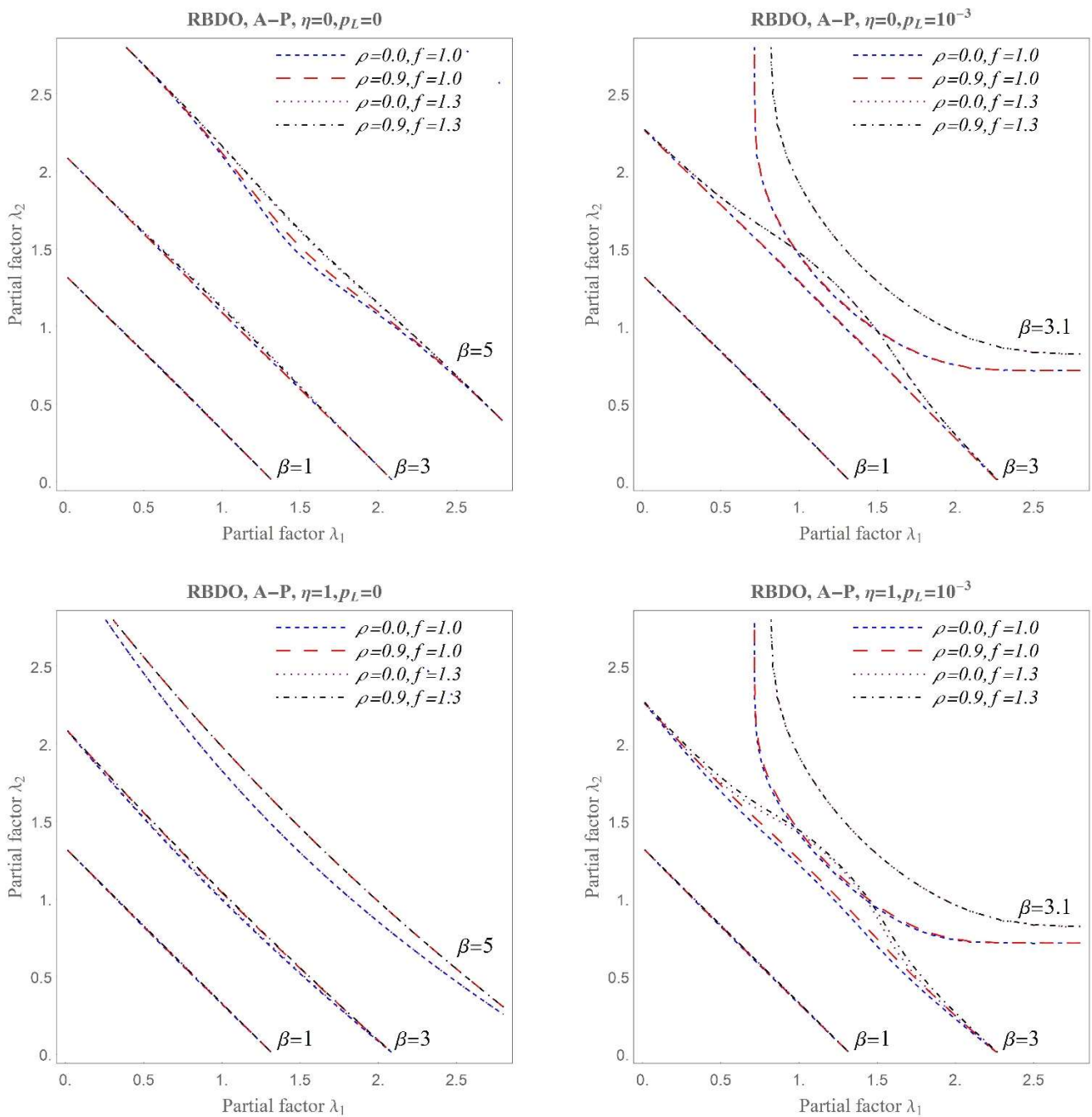

Figure 5: RBDO results for active-passive system with two bars of same material strength.

## Results for Risk Optimization (RO)

Results for risk optimization of the active-passive system are presented in Table 2, and in Figures 6 (fragile material, $\eta = 0$) and 7 (ductile material, $\eta = 1$). By comparing Figures 6 and 7, it becomes clear that material post-failure behavior has a smaller influence in optimal system configuration, as compared to the combined effects of latent probability and impact factors. These figures show contour plots of the objective function for different configurations of the problem. Contour curves grow very fast for small values of partial factors $\lambda_1$ and $\lambda_2$ (bottom left corner in the plots). The objective function is linear for small values of these parameters, as observed for small target reliabilities in RBDO (Figure 5). For $p_L = 10^{-3}$ or smaller (left in Figures 6 and 7), objective functions also grow for large $\lambda_1$ and $\lambda_2$; the upper right corner in these plots is dominated by cost of conservativeness ($\lambda$'s greater than necessary). For $p_L = 10^{-3}$ or smaller, there is an economical trade-off between alternative designs along the diagonal line $\lambda_2 \approx 2.2 - \lambda_1$, where increases in $\lambda_1$ are compensated by reductions in $\lambda_2$. Along this line, however, the cost function varies in a non-linear way, leading to the optimal point $\lambda_1 = \lambda_2 \approx 1.1$, for $f_i = 1$, as shown. For a load amplification factor $f_i = 1.3$, the optimal designs are given by two anti-symmetric solutions with two bars of different areas ($\lambda_1 = 0.512$ and $\lambda_2 = 1.717$, or vice-versa, for $\eta = 0$; $\lambda_1 = 0.565$ and $\lambda_2 = 1.651$, or vice-versa, for $\eta = 1$). The non-linear behavior of the cost function along line $\lambda_2 \approx 2.2 - \lambda_1$ can be associated to expected progressive failure costs ($PC$) in Eq. (31).

When material strengths and COVs are the same, the optimal solutions are symmetric w.r.t. the diagonal line $\lambda_1 = \lambda_2$, i.e., optimal solutions do not differentiate between bars. The sum of optimal partial factors ($\lambda_1 + \lambda_2$) varies around 2.2; showing that most system safety is coming from existence of two elements to share the load. The traditional load factor for these designs varies around $\lambda_E \approx 2.2$. The solutions are similar for fragile and ductile materials, but the conditional reliabilities are quite different: for the ductile material, conditional reliability ($\beta_{2|1}$) is much larger; hence, expected costs for progressive collapse ($PC$) are smaller.

When the optimal system reliability is larger than the latent reliability (right in Figures 6 and 7, for $p_L = 10^{-2}$), the local minimum along diagonal line $\lambda_1 = \lambda_2$ is dislocated towards the upper right corner, and optimal load factors are increased. For $f_i = 1$, the optimal structure has two bars with same cross-section area ($\lambda_1 = \lambda_2 \approx 1.46$). For $f_i = 1.3$, three local minima can be clearly seen (bottom right in Figures 6 and 7); the global minimum solution, however, has two bars with same cross-section area, with $\lambda_1 = \lambda_2 \approx 1.8$. Again, we notice little influence of material post-failure behavior in the optimal solutions.

Usually, structural design is made on a member-by-member basis, and global (system) checks are made considering redundancy, robustness, etc. The problem herein was formulated in such a way that the optimizer decides if a second (redundant) member is required. System checks are warranted by collapse probabilities, and member checks are considered by failure of hyperstatic bars (service failure in Eq. 31). Therefore, some unexpected configurations are obtained, such as for $p_L = 10^{-3}$ and $f_i = 1.3$ in Table 2. The optimal solution contains a bar which cannot sustain the total load ($\lambda_1 = 0.512$ for $\eta = 0$), but which will share the load with bar 2 until it (eventually) fails. This optimal configuration is independent of material model, as for $\eta = 1$ a similar solution is obtained. This feature is clearly a consequence of dynamic amplification in load redistribution ($f_i = 1.3$).

Table 2: RO results for active-passive system with two bars of same material strength ($\mu_1 = \mu_2$).

| | | Latent prob. ($p_L = 10^{-3}$) | | | | Latent prob. ($p_L = 10^{-2}$) | | | |
|---|---|---|---|---|---|---|---|---|---|
| **Param.** | **Correl. ($\rho_{12}$)** | 0.0 | | | | 0.0 | | | |
| | **Material ($\eta$)** | 0 | | 1 | | 0 | | 1 | |
| | **Impact ($f_i$)** | 1.0 | 1.3 | 1.0 | 1.3 | 1.0 | 1.3 | 1.0 | 1.3 |
| **Optimu** | $\lambda_1$ | 1.110 | 0.512 | 1.104 | 0.565 | 1.461 | 1.767 | 1.461 | 1.767 |
| | $\lambda_2$ | 1.110 | 1.717 | 1.104 | 1.651 | 1.461 | 1.767 | 1.461 | 1.767 |
| | $a_1$ | 2.220 | 1.024 | 2.207 | 1.130 | 2.922 | 3.534 | 2.922 | 3.534 |
| | $a_2$ | 2.220 | 3.434 | 2.207 | 3.302 | 2.922 | 3.534 | 2.922 | 3.534 |
| **Reliabilit** | $\beta_1$ | 3.268 | 3.289 | 3.242 | 3.260 | 4.589 | 5.466 | 4.589 | 5.466 |
| | $\beta_{2|1}$ | 0.343 | 0.979 | 3.570 | 2.144 | 1.381 | 1.091 | 5.276 | 4.823 |
| | $\beta_{1\cap2}$ | 3.602 | 3.518 | 3.570 | 3.503 | 5.276 | 6.490 | 5.276 | 6.490 |
| | $\beta_{SYS}$ | 2.942 | 2.925 | 2.991 | 2.975 | 2.719 | 2.741 | 2.719 | 2.741 |
| **Costs** | Material | 1.110 | 1.115 | 1.104 | 1.108 | 1.461 | 1.767 | 1.461 | 1.767 |
| | SF | 0.003 | 0.002 | 0.004 | 0.003 | 0 | 0 | 0 | 0 |
| | PC | 0.015 | 0.007 | 0.009 | 0.001 | 0.030 | 0.005 | 0.030 | 0.005 |
| | DC | 0.105 | 0.157 | 0.110 | 0.163 | 0.175 | 0.283 | 0.176 | 0.283 |
| | Total | 1.232 | 1.281 | 1.227 | 1.275 | 1.667 | 2.054 | 1.667 | 2.054 |

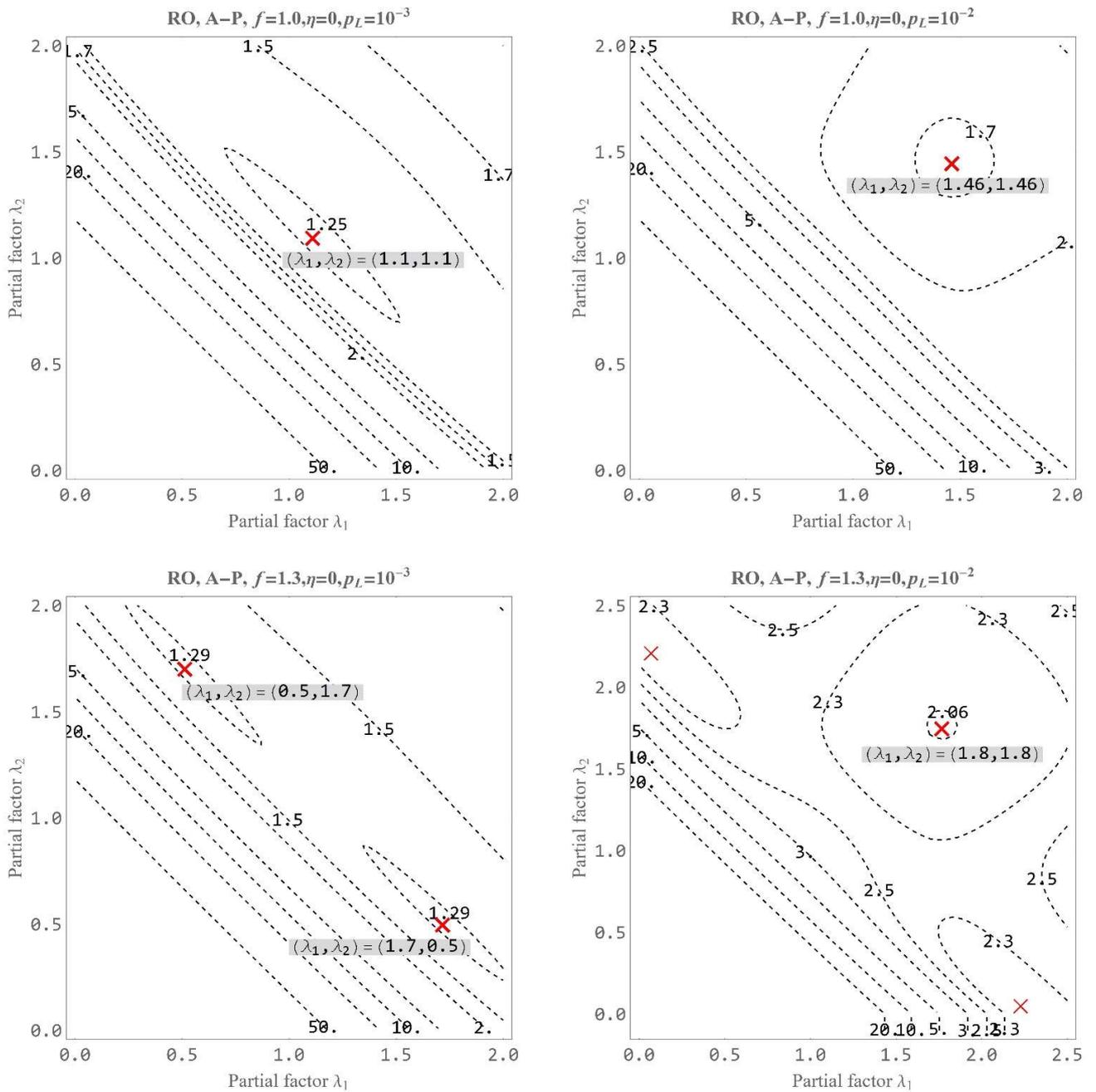

Figure 6: RO results for active-passive system with two bars of fragile material of same strength.

Figure 7: RO results for active-passive system with two bars of ductile material of same strength.

## 6.3 Results for two bars of different mean strength ($\mu_2 = 9\mu_1$)

**Results for System RBDO**

By using system reliability as a design constraint (System RBDO), the results illustrated in Figure 8 are obtained for the case with different mean strengths. Results for fragile material ($\eta = 0$) are shown at the top, and for the ductile material ($\eta = 1$) at the bottom. Results neglecting latent failure probability ($p_L = 0$) are shown left in Figure 8, and results for $p_L = 10^{-3}$ are shown right.

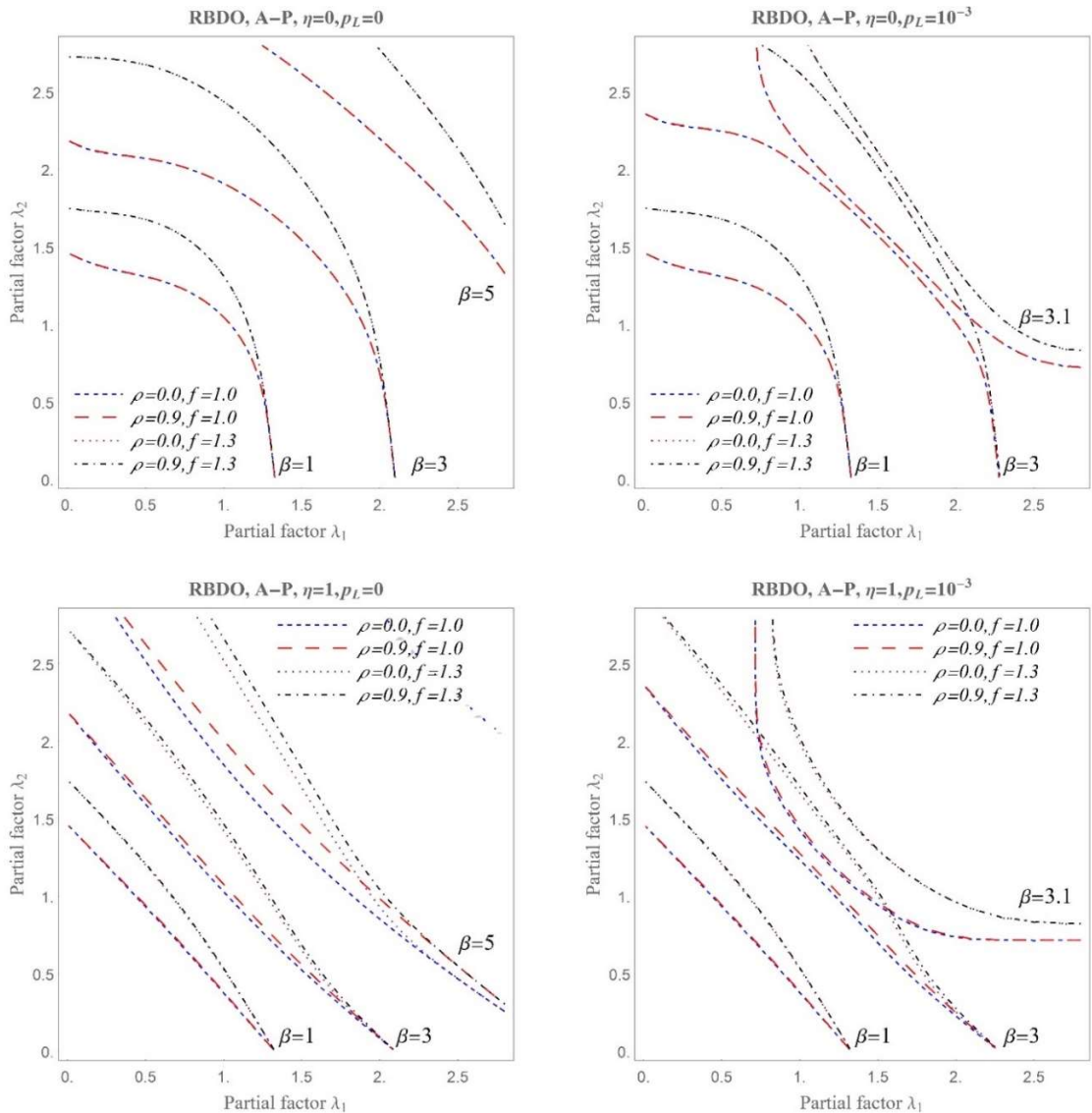

Figure 8: System RBDO results for active-passive system with two bars of different mean material strength.

In comparison to the same strength case, we notice a lack of symmetry (as expected), and a greater influence of material post-failure behavior and load amplification factor. The impact factor has a greater role when $\lambda_1$ is small and $\lambda_2$ is large. System reliability is closer to linear with $\lambda_1$ and $\lambda_2$ for the ductile material with $\beta_T < \beta_L = 3.09$, and is non-linear with $\lambda_1$ and $\lambda_2$ for the fragile material. Linear or non-linear, there is an equivalence between two-bar and single bar solutions, when $\beta_T < \beta_L = 3.09$. For $\beta_T > \beta_L$, only redundant two-bar solutions are optimal. Also for this case, the reliability of the system can only be larger than the latent reliability if the system becomes redundant.

Impact of strength correlation is only relevant for the ductile material and for large $\beta_T$, when latent probability is neglected or when it is much smaller than $\beta_T$.

**Results for Risk Optimization (RO)**

Results for risk optimization of the active-passive system, with different mean strengths, are presented in Table 3 and Figures 9 ($\eta = 0$) and 10 ($\eta = 1$). In comparison to Figures 6 and 7, it is noted that symmetry with respect to diagonal line $\lambda_1 = \lambda_2$ is gone, as expected. The general trend, however, is still there: for $p_L = 10^{-3}$ (left in Figures 9 and 10), there are two local minima along the diagonal line $\lambda_2 \approx$ (2.2 to 2.5) $- \lambda_1$, where a tradeoff between cross-section areas of the two bars is observed; for $p_L = 10^{-2}$ (right in Figures 9 and 10), the global minimum appears at the upper right corner, and both bars become necessary. For $p_L = 10^{-3}$ or smaller, the optimal solutions are statically determinate (or almost): for $f_i = 1$, we have a single bar of the strongest material; for $f_i = 1.3$ the optimal solution has a strong bar of the weakest material (large $\lambda_1$), and a thin wire of the best material (very small $\lambda_2$). For $p_L = 10^{-2}$, optimal solutions become clearly redundant systems, with equivalent load factors. Optimal solutions have bars of nearly inversely proportional cross-section areas, with $a_1\mu_1 \approx a_2\mu_2$.

Two of the results in Table 3 have a vanishing bar 1, with $\lambda_1 = 0$. The corresponding reliability index is $\beta_1 = -2.503$. For this situation, the event $F_1$ "failure of bar 1" is not defined, and Eqs. (26) and (31) could be wrong. However, we observe that objective functions are continuous for very small $\lambda_1$, around the optimal solutions (see Figures 9 and 10, top left plot). Hence, since $F_1$ exists and Eqs. (26) and (31) are correct for very small $\lambda_1$, we conclude that the solutions are correct also for $\lambda_1 = 0$.

Table 3: Risk optimization results for different mean strengths ($\mu_2 = 9\mu_1$).

| | | Latent prob. ($p_L = 10^{-3}$) | | | | Latent prob. ($p_L = 10^{-2}$) | | | |
|---|---|---|---|---|---|---|---|---|---|
| **Param.** | **Correl. ($\rho_{12}$)** | 0.0 | | | | 0.0 | | | |
| | **Material ($\eta$)** | 0 | | 1 | | 0 | | 1 | |
| | **Impact ($f_i$)** | 1.0 | 1.3 | 1.0 | 1.3 | 1.0 | 1.3 | 1.0 | 1.3 |
| **Optimum** | $\lambda_1$ | 0 | 2.213 | 0 | 2.074 | 1.701 | 1.859 | 1.683 | 1.844 |
| | $\lambda_2$ | 2.219 | 0.072 | 2.219 | 0.195 | 1.504 | 1.787 | 1.481 | 1.776 |
| | $a_1$ | 0 | 22.131 | 0 | 20.741 | 17.009 | 18.592 | 16.827 | 18.444 |
| | $a_2$ | 2.465 | 0.081 | 2.465 | 0.217 | 1.671 | 1.986 | 1.646 | 1.973 |
| **Reliability** | $\beta_1$ | -2.503 | 3.271 | -2.503 | 2.994 | 2.456 | 2.908 | 2.405 | 2.871 |
| | $\beta_{2|1}$ | 3.266 | -3.147 | 3.266 | 2.193 | 1.501 | 1.136 | 5.778 | 4.973 |
| | $\beta_{1\cap 2}$ | 3.266 | 3.448 | 3.266 | 3.476 | 5.860 | 6.690 | 5.778 | 6.644 |
| | $\beta_{SYS}$ | 2.865 | 2.957 | 2.865 | 3.009 | 2.841 | 2.778 | 2.879 | 2.791 |
| **Costs** | Material | 1.109 | 1.143 | 1.109 | 1.135 | 1.602 | 1.823 | 1.582 | 1.810 |
| | SF | 0 | 0.001 | 0 | 0.009 | 0.045 | 0.012 | 0.053 | 0.015 |
| | PC | 0.011 | 0.005 | 0.011 | 0 | 0.026 | 0.008 | 0.018 | 0.004 |
| | DC | 0.155 | 0.158 | 0.155 | 0.155 | 0.097 | 0.234 | 0.108 | 0.245 |
| | Total | 1.275 | 1.307 | 1.275 | 1.299 | 1.770 | 2.077 | 1.761 | 2.073 |

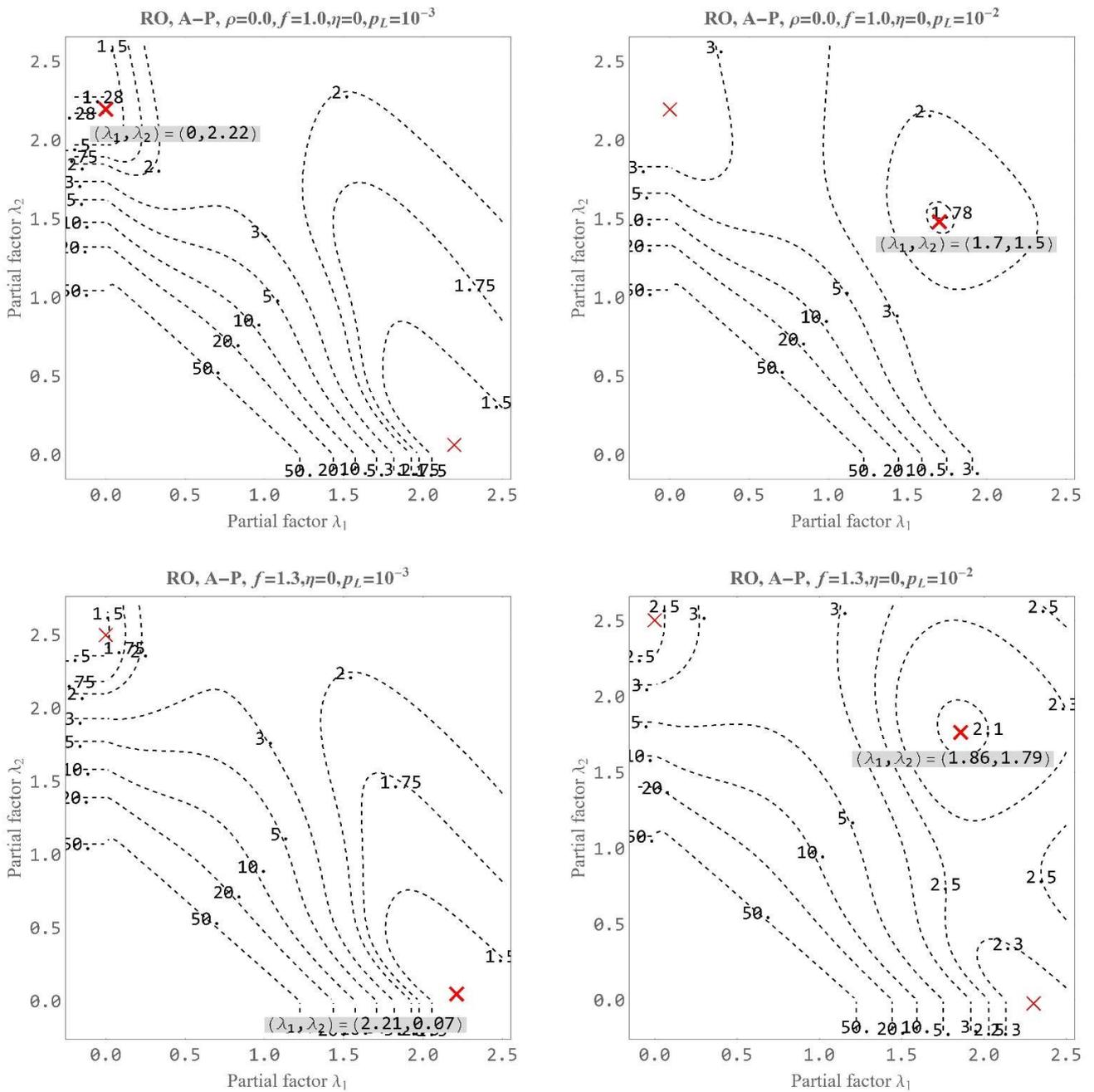

Figure 9: RO results for fragile materials of different mean strengths ($\mu_2 = 9\mu_1$).

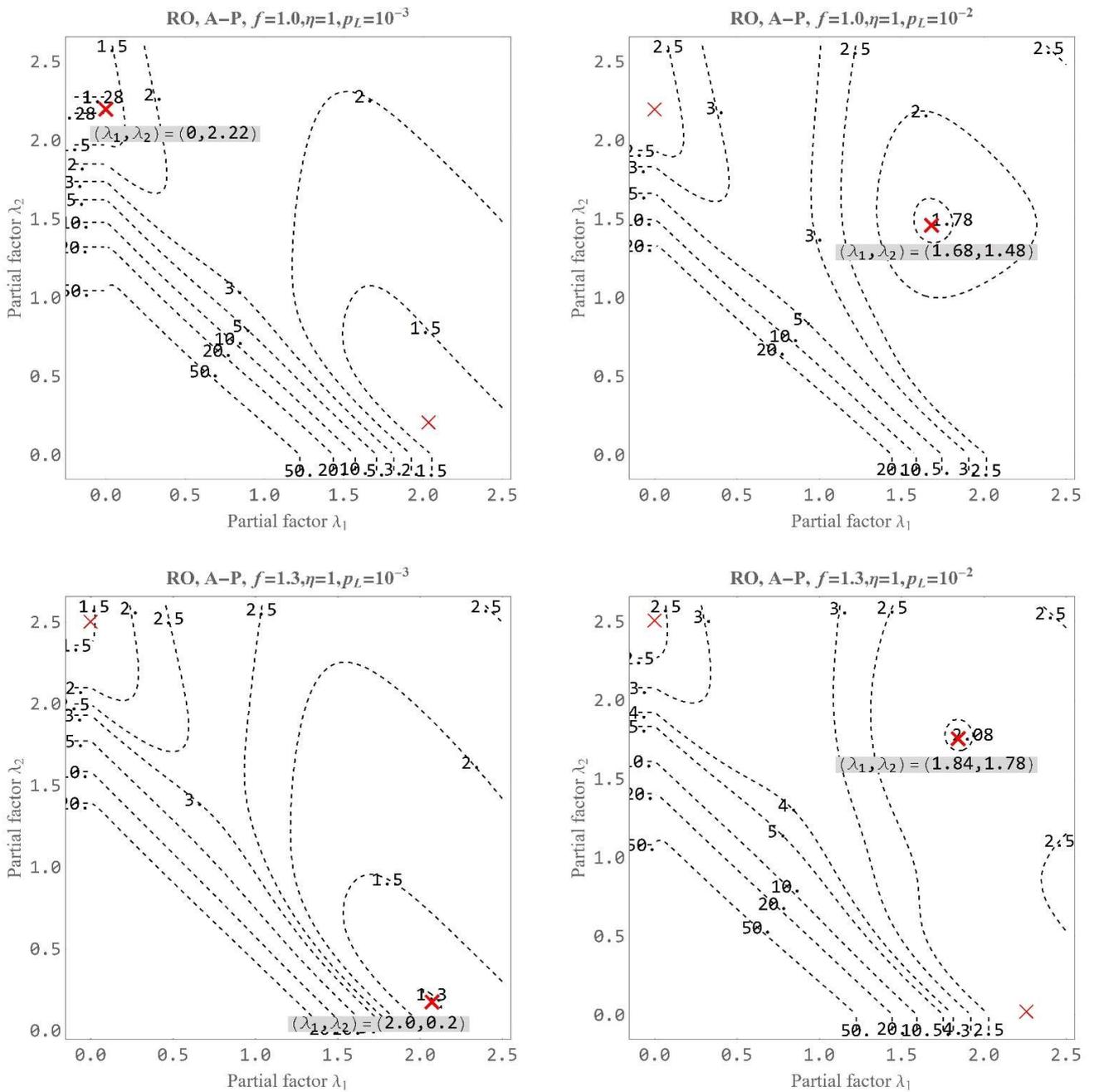

Figure 10: RO results for ductile materials of different mean strengths ($\mu_2 = 9\mu_1$).

## 6.4 Results for two bars of different mean and COV ($\mu_2 = 9\mu_1$; $\delta_1 = 3\delta_2 = 0.15$)

Results for System RBDO are similar to Example 6.3; hence, not presented nor further discussed herein. Also, level curves for the objective functions in risk optimization are similar to Figures 9 and 10; hence, these figures are not presented for the current case.

Results for risk optimization of the active-passive system, with different mean strengths and different COVs, are presented in Table 4. Results are quite similar to those obtained for different means but same COV (Section 5.2): for $p_L = 10^{-2}$, all optimal solutions are redundant two-bar systems; for $p_L = 10^{-3}$, all optimal solutions are single-bar statically determinate systems. The most significant difference is observed for $p_L = 10^{-3}$ and for $f_i = 1.3$, where the optimal solution changed from a two-bar solution with a thin wire as bar two, to a single bar solution made of material 2 only. In all cases for which bar one has zero area, the joint failure reliabilities converge to that of bar two: $\beta_{1\cap 2} = \beta_2$.

Table 4: Risk optimization results for materials of different mean strength and different COV.

| | | Latent prob. ($p_L = 10^{-3}$) | | | | Latent prob. ($p_L = 10^{-2}$) | | | |
|---|---|---|---|---|---|---|---|---|---|
| **Param.** | **Correl. ($\rho_{12}$)** | 0.0 | | | | 0.0 | | | |
| | **Material ($\eta$)** | 0 | | 1 | | 0 | | 1 | |
| | **Impact ($f_i$)** | 1.0 | 1.3 | 1.0 | 1.3 | 1.0 | 1.3 | 1.0 | 1.3 |
| **Optimum** | $\lambda_1$ | 0 | 0 | 0 | 0 | 1.796 | 1.940 | 1.775 | 1.917 |
| | $\lambda_2$ | 2.067 | 2.401 | 2.067 | 2.401 | 1.502 | 1.807 | 1.474 | 1.783 |
| | $a_1$ | 0 | 0 | 0 | 0 | 17.957 | 19.395 | 17.754 | 19.171 |
| | $a_2$ | 2.297 | 2.668 | 2.297 | 2.668 | 1.669 | 2.007 | 1.637 | 1.981 |
| **Reliability** | $\beta_1$ | -2.551 | -2.423 | -2.551 | -2.423 | 2.290 | 2.595 | 2.247 | 2.554 |
| | $\beta_{2\|1}$ | 3.363 | 2.699 | 3.363 | 2.699 | 1.624 | 1.265 | 5.514 | 4.871 |
| | $\beta_{1\cap 2}$ | 3.363 | 4.336 | 3.363 | 4.336 | 5.603 | 6.423 | 5.514 | 6.353 |
| | $\beta_{SYS}$ | 2.917 | 2.615 | 2.917 | 2.615 | 2.856 | 2.784 | 2.908 | 2.819 |
| **Costs** | Material | 1.034 | 1.201 | 1.034 | 1.201 | 1.649 | 1.873 | 1.625 | 1.850 |
| | SF | 0 | 0 | 0 | 0 | 0.075 | 0.034 | 0.085 | 0.040 |
| | PC | 0.008 | 0.069 | 0.008 | 0.069 | 0.026 | 0.013 | 0.017 | 0.004 |
| | DC | 0.139 | 0.101 | 0.139 | 0.101 | 0.086 | 0.205 | 0.098 | 0.223 |
| | Total | 1.180 | 1.371 | 1.180 | 1.371 | 1.835 | 2.125 | 1.825 | 2.116 |

## 7. CONCLUDING REMARKS

In this paper, the optimal design of redundant systems was addressed. Progressive collapse was addressed in an objective way, differentiating the consequences of direct collapse, for statically determinate structures, and progressive collapse, for redundant statically indeterminate structures. The study addressed System Reliability-Based Design Optimization (System RBDO) and consequence-driven or Risk-based Optimization (RO). A comprehensive study was performed, considering the effects of material post-failure behavior, strength correlation, dynamic amplification in load redistribution, different ratios of mean and variance of material strengths, and the impact of non-structural factors, or factors beyond structural design.

Among all parameters studied, it was shown that non-structural factors, or factors beyond structural design, play the largest role in optimal configurations. These factors include epistemic uncertainties affecting abnormal loadings; human errors in design, construction, and operation; operational abuse; etc. The influence of such factors in the optimum structural design problem was considered by imposing a latent failure probability, resulting from the encompassing risk analysis. It was shown that the only way to make system failure probability smaller than the latent failure probability is by making the structural system redundant. Hence, as a general practical conclusion, structural systems need to be redundant not (only) because of known, aleatory uncertainties, but also because of epistemic uncertainty arising from non-structural factors. Drawing a parallel to the discussion in Chapter 13 of ref. [46], it may not be worth increasing safety factors to account for gross errors, but it is worth making structures redundant because of them, as shown herein. Results in this paper where obtained for very simple two-bar parallel systems, with epistemic uncertainty affecting the reliability of connections. However, the results motivate intuitive generalization to the optimal design of more complex redundant systems.

It is well known that, in order to increase system reliability using shelf components, one has to create redundancy using parallel components. However, in structural engineering, each component is designed individually. Hence, one can in principle design individual members to be as reliable as desired. But when non-structural factors are taken into account, it becomes evident that redundancy is required. This is the main conclusion of this study. Somehow, this has been overlooked in past research addressing structural optimization via system RBDO or risk optimization [19-43]. Specifically w.r.t these two formulations, it was shown herein that when the latent reliability is larger than target system reliability in RBDO, or larger than optimal risk-based reliabilities, there is an equivalence between redundant and non-redundant designs, and results become dependent on impact factors. However, when target (System RBDO) or optimal reliabilities are larger than the latent reliability, optimal designs become necessarily redundant. Optimal designs were found largely independent of material post-failure behavior (ductile-fragile), but only because same failure cost multipliers were considered. Brittle failures are known to have worst consequences, because of the absence of warning prior to failure. Strength correlation was shown to play an insignificant role in optimal design. Bars with same mean strength and COV resulted in symmetrical optimal designs, whereas for different mean and COV ratios, symmetry is lost.

Ongoing research is addressing the impacts of latent failure probabilities in the optimal design of steel trusses and RC frames. Topology optimization concepts are being employed to handle the multi-member problem. Understanding the fundamentals of reliability of structural systems in progressive collapse is very important, before more practical design situations can be considered.

## ACKNOWLEDGEMENTS

Fruitful discussions of an earlier version of this manuscript with Prof. André Jacomel Torii is deeply acknowledged. Funding of this research project by Brazilian agencies CAPES (Brazilian Higher Education Council), CNPq (Brazilian National Council for Research, grant n. 306373/2016-5) and FAPESP (São Paulo State Foundation for Research, grant n. 2017/01243-5) is also acknowledged.

## CONFLICT OF INTEREST

The author of this manuscript states that there is no conflict of interest.

## REPLICATION OF RESULTS

The Mathematica algorithms used to obtain results presented in this paper will be made available upon request.

**APPENDIX: RESULTS FOR SYSTEM WITH PASSIVE (STANDBY) REDUNDANCY**

This paper addresses design of redundant systems considering progressive collapse. Several problem parameters are considered, such as of material post-failure behavior, strength correlation, dynamic load amplification and ratios of mean and variance of material strengths. Surely, the type of redundancy is also relevant. In the body of the manuscript we addressed conventional systems, where redundancy is active, but may become passive in case of failure of a primary element. In this Appendix, we address passive (standby) redundancy, as illustrated in Figure 2 (right). All symbols have the same meaning as in the manuscript body, with exception of the design variables, which are renamed from $\lambda_1$ to $\lambda_A$ (active member), and from $\lambda_2$ to $\lambda_P$ (passive member). Notice that the problem is no longer symmetric.

**Formulation**

Figure 11 illustrates the event three for a passive two-bar system submitted to two load applications. When the structure is first loaded, only bar 1 is active. If the connection to bar 1 fails, bar 2 and its connection are automatically mobilized. If connection to bar two also fails, we have direct collapse (*DC*). If the connection to bar 2 does not fail, the strength of the second bar is mobilized. If the connection to bar 1 does not fail, uppon first load appplication, then the chain of events ilustrated in the upper branch in Figure 11 becomes possible.

System failure is characterized by failure of the redundant element; which can occur immediately following failure of the primary element, or in a subsequent (independent) load application. Limit state and reliability index for the basic events ($F_1$, $F_{2|1}$) are those of Section 4.4 (Eqs. 17 to 20).

The standby element may be engaged before failure of element 1; for instance, due to excessive deformation of element 1. If this is the case, then both elements can also fail simultaneously ($F_{1\cap 2}$, path "i" in Figure 11). Dynamic load amplification may or may not occur, in this case. If dynamic load amplification is considered, then $f_i$ must be included in Eqs. (21) and (22). However, if $\lambda_2$ goes to zero in the optimization problem ($\lambda_2 \to 0$), with $f_i > 1$; then it is possible that $\Phi[-\beta_{1\cap 2}]$ becomes greater than $\Phi[-\beta_1]$, following Eqs. (18) and (22). In this case, the probability of the event 'only bar one fails' should go to zero (as bar two would fail also); for this reason, the operator $\max[0, \ldots]$ is included in Eq. (25):

$$P[F_1 \text{ only}] = \max[0, \Phi[-\beta_1] - \Phi[-\beta_{1\cap 2}]]. \tag{33}$$

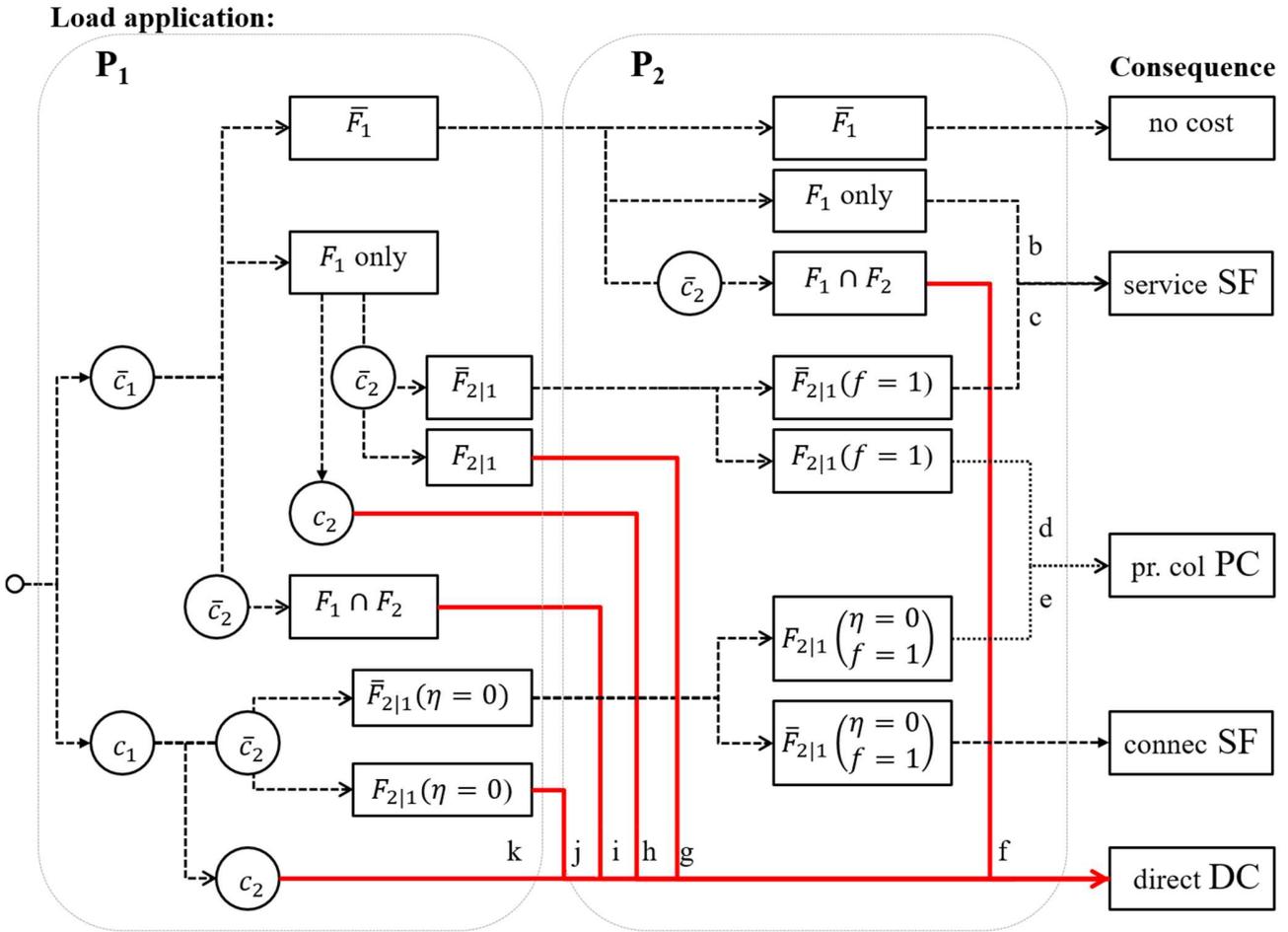

Figure 11: Event tree for passive two-bar system submitted to two load applications.

If the standby element does not engage before failure of element 1, then $P[F_{1 \cap 2}] = 0$, and path "i" in Figure 11 and line (i) in equation (34) should be disregarded.

Following the reasoning in Section 4.4 and Figure 11, the system failure probability is obtained as:

$$p_{fsys}(\lambda_A, \lambda_P) = (1 - p_1)(1 - p_2)(P[F_1 \text{ only}])\Phi[-\beta_{2|1}]\Phi[-\beta_{2|1}(f_i = 1)] \quad \text{(d)}$$

$$+(p_1)(1 - p_2)(1 - \Phi[-\beta_{2|1}(\eta = 0)])\Phi[-\beta_{2|1}(\eta = 0, f_i = 1)] \quad \text{(e)}$$

$$+(1 - p_1)(1 - p_2)(1 - \Phi[-\beta_1])\Phi[-\beta_{1 \cap 2}] \quad \text{(f)}$$

$$+(1 - p_1)(1 - p_2)(P[F_1 \text{ only}])\Phi[-\beta_{2|1}] \quad \text{(g)}$$

$$+(1 - p_1)(p_2)(P[F_1 \text{ only}]) \quad \text{(h)}$$

$$+(1 - p_1)(1 - p_2)\Phi[-\beta_{1 \cap 2}] \quad \text{(i)}$$

$$+(p_1)(1-p_2)\Phi[-\beta_{2|1}(\eta=0)] \tag{j}$$

$$+(p_1)(p_2) \tag{k}$$

$$\tag{34}$$

The letters identifying Equation lines (34 d-k) correspond to the paths to collapse illustrated in Figure 11. Eq. (34) assumes independence between the connection failure events $c_1$ and $c_2$.

For the passive redundant system, and following the reasoning in Section 5.2, the objective function of risk optimization is obtained as:

$$h_{RO}(\mathbf{d}) = \frac{(a_1(\lambda_A)\mu_1 + a_2(\lambda_P, f_i)\mu_2)}{(a_1(1)\mu_1 + a_2(1,1)\mu_2)} \tag{a}$$

$$+k_{SF}\frac{a_1(\lambda_A)}{a_1(1)}(1-p_1)(1-\Phi[-\beta_1])(P[F_1 \text{ only}]) \tag{b}$$

$$+k_{SF}\frac{a_1(\lambda_A)}{a_1(1)}(1-p_1)(1-p_2)(P[F_1 \text{ only}])$$

$$(1-\Phi[-\beta_{2|1}])(1-\Phi[-\beta_{2|1}(f=1)]) \tag{c}$$

$$+k_{PC}(1-p_1)(1-p_2)(P[F_1 \text{ only}])(1-\Phi[-\beta_{2|1}])\Phi[-\beta_{2|1}(f=1)] \tag{d}$$

$$+k_{PC}(p_1)(1-p_2)(1-\Phi[-\beta_{2|1}(\eta=0)])\Phi[-\beta_{2|1}(\eta=0,f=1)] \tag{e}$$

$$+k_{DC}(1-p_1)(1-p_2)(1-\Phi[-\beta_1])\Phi[-\beta_{1\cap2}] \tag{f}$$

$$+k_{DC}(1-p_1)(1-p_2)(P[F_1 \text{ only}])\Phi[-\beta_{2|1}] \tag{g}$$

$$+k_{DC}(1-p_1)(p_2)(P[F_1 \text{ only}]) \tag{h}$$

$$+k_{DC}(1-p_1)(1-p_2)\Phi[-\beta_{1\cap2}] \tag{i}$$

$$+k_{DC}(p_1)(1-p_2)\Phi[-\beta_{2|1}(\eta=0)] \tag{j}$$

$$+k_{DC}(p_1)(p_2) \tag{k}$$

$$\tag{35}$$

Notice in Figure 11 that there are two failure paths which lead to service failure; this explains the second line in Eq. (35).

**Results for System RBDO, two bars of same mean strength ($\mu_1 = \mu_2$)**

Results of System RBDO are illustrated in Figure 12, for the passive assembly with same mean material strengths. Results for fragile material ($\eta = 0$) are shown at the top, and for the ductile material ($\eta = 1$) at the bottom. Results neglecting latent failure probability ($p_L = 0$) are shown left in Figure 12, and results for $p_L = 10^{-3}$ are shown right. Note the relevant impact of the latent failure probability, as observed for the active-passive system in Figure 5.

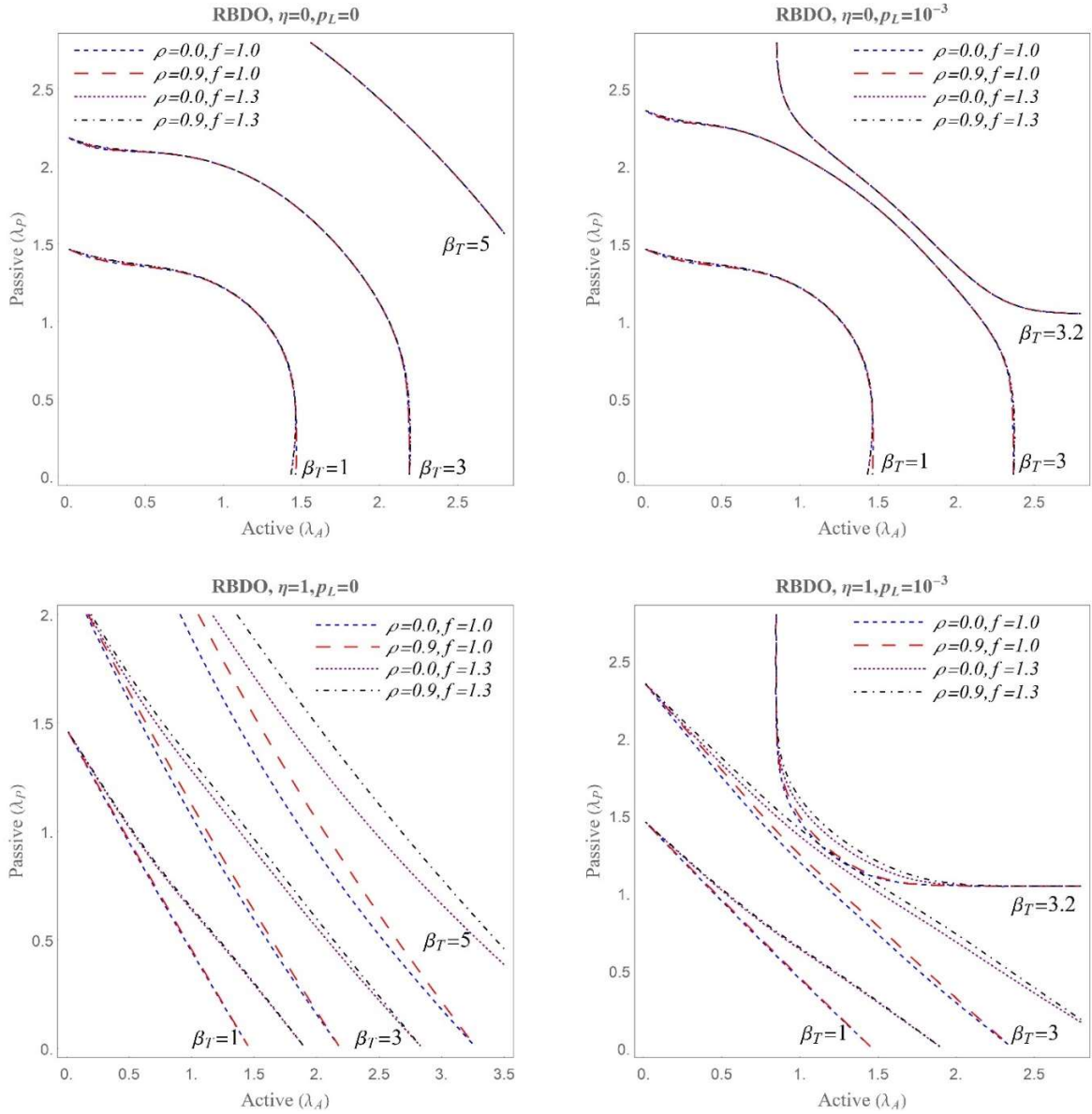

Figure 12: System RBDO results for passive system with two bars of same material strength.

Comparing the results in Figure 12 with those in Figure 5, for an active-passive system with similar characteristics, one finds out that material post-failure behavior has a greater influence in optimal design of passive systems. Results for fragile material ($\eta = 0$, top in Figure 5) are non-linear with design factors $\lambda_A$ and $\lambda_P$, and insensitive to dynamic load amplification factor $f_i$ and strength correlation $\rho_{12}$, regardless of latent failure probability. Results for ductile material ($\eta = 1$, bottom in Figure 12) show greater influence of $f_i$ and $\rho_{12}$, when $\beta_T < \beta_L$. For the ductile material, System RBDO results are almost linear with design factors $\lambda_A$ and $\lambda_P$ for $\beta_T < \beta_L$, similar to the active-passive system.

**Results for RO, two bars of same mean strength ($\mu_1 = \mu_2$)**

Results for risk optimization of the passive system are presented in Table 6, and in Figures 13 (fragile material, $\eta = 0$) and 14 (ductile material, $\eta = 1$). These figures show contour plots of the objective function for different configurations of the problem. Contour curves grow very fast for small values of partial factors $\lambda_A$ and $\lambda_P$ (bottom left corner in the plots), as these are unconservative designs. For the passive system, the objective functions are not symmetrical w.r.t the line $\lambda_A = \lambda_P$, since the system is not symmetrical either (bar 2 remains in standby until required). In comparison to the equivalent active-passive system (Figures 6 and 7), objective functions are more non-linear with parameters $\lambda_A$ and $\lambda_P$, and material post-failure behavior has a greater impact. Both behaviors were also identified for the System RBDO formulation (see Figures 5 and 12).

For $p_L = 10^{-3}$ or smaller (left in Figures 13 and 14), two local minima can be observed; one for large $\lambda_A$ and very small (or even zero $\lambda_P$), one for $\lambda_A = 0$ and large $\lambda_P$. There is an economical trade-off between these alternative designs, as objective functions are similar. The global minimum, in all cases, is the design with a large primary element (large $\lambda_A$) and very small secondary element (very small $\lambda_P$). For brittle material, the second bar is not strong enough to sustain mean loads, and conditional reliabilities are very low ($\beta_{2|1} < 0$). Because the ductile material continues to take load after failure, conditional reliabilities are large, in spite of small area for the second bar.

When $p_L = 10^{-2}$ (right in Figures 13 and 14), a third minimum point appears, and becomes the global minimum: in all cases, this corresponds to a more robust redundant systems, having two bars with similar cross-section areas. Interestingly, the optimal solutions are the same for both materials, showing that redundancy is mostly required to overcome the large connection failure probability. Since connection failure is controlling optimal design, material post-failure behavior becomes irrelevant. Also for $p_L = 10^{-2}$ we note that optimal solutions are the same, regardless of dynamic load amplification factor. This may be a consequence that the redundant element is designed already considering the amplification factor (in contrast to Eq. 15). Connection reliability has a strong impact in direct collapse costs, as observed in Table 5.

Table 5: RO results for passive system with two bars of same mean material strength ($\mu_1 = \mu_2$).

| | | **Latent prob. ($p = 10^{-3}$)** | | | | **Latent prob. ($p = 10^{-2}$)** | | | |
|---|---|---|---|---|---|---|---|---|---|
| **Param.** | **Correl. ($\rho_{12}$)** | 0.0 | | | | 0.0 | | | |
| | **Material ($\eta$)** | 0 | | 1 | | 0 | | 1 | |
| | **Impact ($f_i$)** | 1.0 | 1.3 | 1.0 | 1.3 | 1.0 | 1.3 | 1.0 | 1.3 |
| **Optimum** | $\lambda_A$ | 2,286 | 2,269 | 2,075 | 2,095 | 1,909 | 1,938 | 1,909 | 1,938 |
| | $\lambda_P$ | 0,000 | 0,435 | 0,192 | 0,568 | 1,554 | 1,481 | 1,554 | 1,481 |
| | $a_1$ | 4,572 | 4,537 | 4,151 | 4,189 | 3,817 | 3,875 | 3,817 | 3,875 |
| | $a_2$ | 0,000 | 1,130 | 0,385 | 1,477 | 3,107 | 3,851 | 3,107 | 3,851 |
| **Reliability** | $\beta_1$ | 3,410 | 3,373 | 2,948 | 2,992 | 2,555 | 2,625 | 2,555 | 2,625 |
| | $\beta_{2\|1}$ | -3,333 | -1,865 | 3,470 | 3,416 | 1,639 | 1,438 | 1,639 | 1,438 |
| | $\beta_{1 \cap 2}$ | 3,410 | 3,373 | 3,470 | 3,416 | 6,346 | 5,383 | 6,346 | 5,383 |
| | $\beta_{SYS}$ | 2,939 | 2,922 | 2,964 | 2,942 | 2,988 | 2,891 | 2,988 | 2,891 |
| **Costs** | Material | 1,143 | 1,417 | 1,134 | 1,417 | 1,731 | 1,931 | 1,731 | 1,931 |
| | SF | 0,000 | 0,000 | 0,011 | 0,009 | 0,038 | 0,031 | 0,038 | 0,031 |
| | PC | 0,000 | 0,001 | 0,000 | 0,001 | 0,015 | 0,020 | 0,015 | 0,020 |
| | DC | 0,265 | 0,271 | 0,251 | 0,255 | 1,072 | 1,101 | 1,072 | 1,101 |
| | Total | 1,408 | 1,688 | 1,396 | 1,682 | 2,855 | 3,083 | 2,855 | 3,083 |

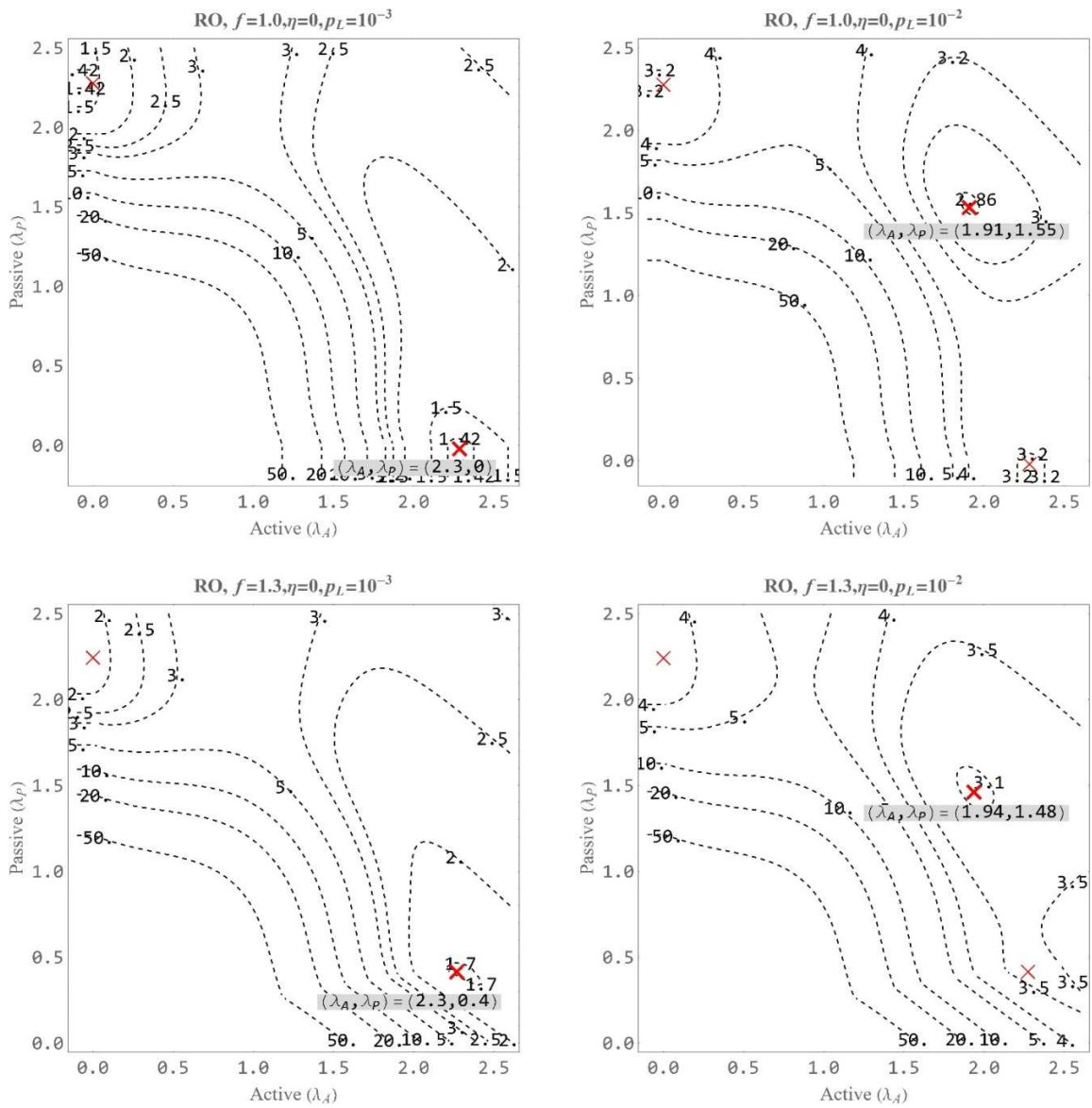

Figure 13: RO results for passive (standby) system with two bars of fragile material of same strength.

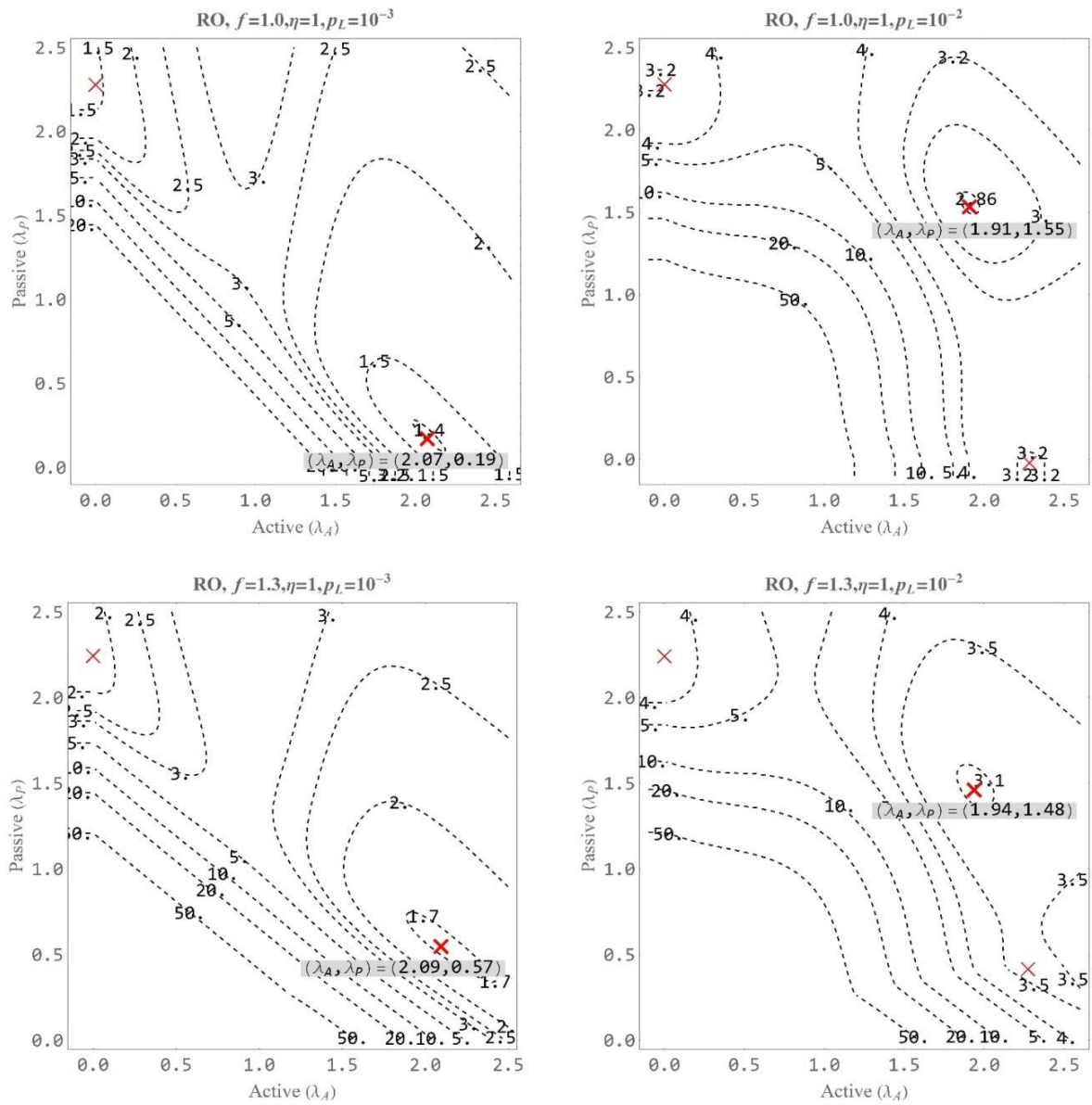

Figure 14: RO results for passive (standby) system with two bars of ductile material of same strength.